\documentclass[10pt, journal]{IEEEtran}
\usepackage[utf8]{inputenc}
\usepackage{cite}
\usepackage{romannum}
\usepackage{amsfonts}
\usepackage{mathtools}
\usepackage{amsmath, bm, amsthm}
\usepackage{amssymb}
\usepackage{float}
\usepackage{graphicx}
\usepackage{caption}
\usepackage{bigints}
\usepackage{subfigure}
\usepackage{cuted}
\usepackage{siunitx}
\usepackage[super]{nth}
\usepackage[table]{xcolor}
\usepackage{orcidlink}
\usepackage{dsfont}
\usepackage{makecell}
\usepackage{xpatch}
\usepackage{bigints}
\hypersetup{hidelinks}
\usepackage{romannum}
\usepackage{tikz}\usetikzlibrary{calc}
\newtheorem{remark}{Remark}

\title{RIS-Assisted Millimeter Wave Communications for Indoor Scenarios: Modeling and Coverage Analysis}
\author{Zhi Chai, \IEEEmembership{Graduate Student Member, IEEE}, Jiajie Xu, \IEEEmembership{Member, IEEE}, Justin P. Coon, \IEEEmembership{Senior Member, IEEE},  Mohamed-Slim Alouini, \IEEEmembership{Fellow, IEEE}
\thanks{Z. Chai, and J. P. Coon are with the Department of Engineering Science, University of Oxford, Parks Road, Oxford, OX1 3PJ, United Kingdom. (e-mail: zhi.chai@eng.ox.ac.uk, justin.coon@eng.ox.ac.uk).\\ \indent J. Xu and M.-S. Alouini are with the Computer, Electrical and Mathematical Sciences and Engineering Division, King Abdullah University of Science and Technology, Thuwal, 23955, Kingdom of Saudi Arabia. (e-mail: jiajie.xu.1@kaust.edu.sa, slim.alouini@kaust.edu.sa).}
}

\date{}
\begin{document}
\pagenumbering{arabic}
\maketitle
\begin{abstract}

\textcolor{black}{Millimeter wave (mmWave) communications and reconfigurable intelligent surfaces (RIS) are two critical technologies for next-generation networks, especially in dense indoor environments. However, existing analyses often oversimplify the indoor environment by neglecting some of the key characteristics, such as height variations, boundary effects, blockage effects, and user spatial distributions. In this paper, we develop an improved stochastic geometry-based model for RIS-assisted mmWave communications in indoor scenarios like conference centers, hospitals, and shopping malls. The proposed model incorporates the height factor for all the nodes in the network (e.g., transmitters, users, RISs, and obstacles) and captures the user clustering behavior in these scenarios. In addition, the boundary effect is also being considered for line-of-sight (LOS) probability calculation. Analytical expressions for distance distributions, LOS probabilities, and the coverage probability (CP) are derived. The CP is then validated through Monte Carlo simulations. Our results reveal deployment insights by approximating and simplifying the derived CP expressions, showing how transmitter density, obstacle density, RIS density, and user cluster radius impact network coverage. Notably, we show that RISs significantly improve coverage when transmitters or transmit power are limited but offer marginal benefits when transmitter density is high. These findings provide practical guidelines for the design and deployment of RIS-assisted indoor mmWave networks.}

\end{abstract}
\section{Introduction}
\subsection{Motivation}

With the depletion of the current commercial spectrum for wireless communications and the soaring demand for connection from consumers, more and more attention is paid to the higher frequency bands such as millimeter wave (mmWave) due to the substantial bandwidth, the fast transmission rate, and the low latency\cite{wang2018millimeter,mezzavilla2018end,10623156}. 
However, a side effect of using mmWave technology is that the signal will experience severe path loss and be hugely attenuated by scatterers, especially in some crowded environments, which leads to limited coverage. Reconfigurable intelligent surface (RIS) technology, as widely studied, is considered one of the most promising solutions to enhance the coverage of wireless communications, especially for high-frequency communications, such as mmWave communications and terahertz communications \cite{10768033,10720781}.
RIS is defined as a synthesized surface that comprises numerous antenna elements with tunable devices (e.g., positive-intrinsic-negative diodes, varactors) mounted on top to manipulate the phase of the impinging waves to realize a constructive superposition after reflection \cite{di2020smart,pei2021ris,10833623}. The reflected signal with an enhanced amplitude can improve the signal-to-noise ratio (SNR) to achieve a broader coverage\cite{yang2020coverage,zeng2020reconfigurable,10663434}.
In \cite{9745078}, a multiple RIS-assisted single-input single-output (SISO) wireless communication system is proposed. The results indicate that the number of RISs, the number of RIS elements, and the communication environment significantly impact network capacity, particularly in large-scale outdoor scenarios.
Authors in \cite{9775205} focus on the influence of local scatterers in real propagation environments within RIS-assisted multiple-input multiple-output (MIMO) communication. They model the space as a three-dimensional cylinder, considering line-of-sight (LoS), single-bounced reflection (SBR) at the RIS, and double-bounced (DB) modes. Stochastic tools are applied to analyze the distance distributions involved in outdoor application scenarios.
In \cite{9576697}, the authors explore free-space RIS-assisted communication in far-field operation, where an arbitrary three-dimensional space is considered. Simulation and analysis results demonstrate that RISs can enhance the performance of mmWave communication. 

\textcolor{black}{It is widely acknowledged that RISs can significantly enhance wireless communication performance, particularly for mmWave and THz systems \cite{chen2024coverage,cui2021snr,10118900}. 
While indoor RIS-assisted mmWave communications have been studied, many existing analyses adopt simplified models that overlook several critical challenges unique to indoor environments. Specifically, current models often:
\begin{itemize}
    \item Neglect the height variations of users, transmitters, RISs, and obstacles \cite{qin2022indoor,li2021enhancing,10279507};
    \item Ignore boundary effects (see Appendix A for details about boundary effects) caused by the finite size of indoor spaces (e.g., assuming RISs are all deployed in the interior of the room) \cite{10279507};
    \item Ignore the association policy when choosing an RIS \cite{li2021enhancing,qin2022indoor,yildirim2020modeling};
    \item Assume overly simplistic network topologies, typically considering only a single transmitter, a single RIS (or two RISs), and a single user \cite{qin2022indoor,yildirim2020modeling};
    \item Neglect blockage effects caused by obstacles or assume the indirect link through RIS is the only viable path\cite{li2021enhancing,10279507,yildirim2020modeling}.
\end{itemize}
These challenges must be addressed to design and analyze RIS-assisted mmWave networks that reflect realistic indoor scenarios. Motivated by these challenges, this paper focuses on developing a more comprehensive and realistic model for RIS-assisted mmWave indoor communications.} The proposed model explicitly accounts for independent random spatial processes of transmitters, RISs, users (end-users, EUs), and obstacles, incorporating factors such as height distributions, obstacle sizes, spatial densities, and user clustering behaviors around indoor landmarks. Our goal is to provide more accurate performance evaluation and practical deployment guidelines for RIS-assisted mmWave indoor networks, targeting environments such as conference centers, hospitals, libraries, and shopping malls.

\subsection{Related Work}
In this section, we provide a brief summary of the related works in RIS-assisted mmWave communication in two general directions of interest to this paper: 1) Stochastic Geometry-Based Analysis of the RIS-Assisted Wireless Communication Network; 2) RISs-Assisted Indoor mmWave Communication.

\subsubsection{Stochastic Geometry-Based Analysis of the RIS-Assisted Wireless Communication Network}
As a passive relay, RIS is widely researched to improve communication performance. In \cite{zhu2020stochastic}, base stations (BSs) and RISs are assumed to follow two independent homogeneous Poisson point processes (HPPPs), respectively. It is also assumed that the density of the BS is lower than the density of the RIS, and there is no LOS path. The association policy proposed in \cite{zhu2020stochastic} follows the shortest distance principle\footnote{The shortest distance principle means the user will associate the nearest RIS in the indirect path.}. Through a comparison with conventional networks, it is concluded that the RIS-assisted network can achieve higher capacity and energy efficiency under the condition that the RIS density is higher than the BS density. It is shown in \cite{zhu2020stochastic} that when the BS density is higher than the RIS density, it is unnecessary to deploy RIS devices. In \cite{shafique2022stochastic}, an application scenario with multiple users, BSs, and RISs in the presence of interference from BSs and RISs is considered, the BSs are assumed to follow an HPPP while the RIS is assumed to follow a binomial point process (BPP), and the Laplace transform of the received signal power, approximated aggregate interference, CP, ergodic capacity, and the energy efficiency are derived based on the assumptions. The analytical results are validated by Monte Carlo simulations. In \cite{zhang2021reconfigurable}, stochastic geometry is used to analyze the performance of a RIS-aided multi-cell non-orthogonal multiple access (NOMA) network. It is shown in \cite{zhang2021reconfigurable} that the achievable data rate reaches an upper limit with the increase in the RIS size and the path loss intercept can be enhanced to improve the network performance without influencing the bandwidth. In \cite{kishk2020exploiting}, the authors propose to coat RIS on a subset of scatterers in the environment, and the light-of-sight (LOS) probability is taken into consideration. The authors model the locations of BSs and users following two independent HPPPs. The scatterers follow the line boolean model proposed in \cite{bai2014analysis}. It is concluded that the deployment of RIS highly improves the coverage regions of the BSs, and it is shown that 6 RIS/$\textrm{km}^2$ is enough to maintain the ratio of blind spots over the total area below $10^{-5}$ when the scatterer density is 300 scatterer/$\textrm{km}^2$. In \cite{10064007}, the spatial correlation between transmitters and RISs is described by using the Gaussian point process (GPP), and the CP expressions under a fixed association strategy and the nearest association strategy are derived in closed form in the presence of interference. It is also shown that the system performance is independent of the density of transmitters with the nearest association strategy when the scenario is interference-limited. In \cite{li2022ris}, researchers analyzed the RIS-assisted mmWave network with an emphasis on whether fewer large RISs or more small RISs can improve the network performance in terms of CP. The model is built in three dimensions (3D) with the height considered. It is concluded that dense small RISs are more favorable than sparse large RISs in places where the obstacle density is high, while sparse large RISs are more favorable in places where the obstacle density is low. However, the heights of the obstacles (e.g., humans, buildings) are constants. In addition, no association policy is introduced related to RIS selection. \textcolor{black}{Besides the coverage analysis of conventional RIS-assisted wireless communication networks, the coverage analysis of RIS-assisted integrated sensing and communication (ISAC) networks, as an emerging topic, has also been investigated recently \cite{10938034}. \cite{10938034} derived the joint CP of ISAC performance with the consideration of the coupling effects of the ISAC dual functions. It showed a 62\% to 97\%  ISAC performance enhancement with the deployment of RISs.} 

\subsubsection{RISs-Assisted Indoor mmWave Communication}
For RIS-assisted indoor mmWave network performance analysis, related work includes \cite{li2021enhancing,qin2022indoor}. \cite{li2021enhancing} discussed small-cell densification and RIS deployment for coverage improvement in a 2-dimensional (2D) room. For the RIS solution, the deployment of RISs was assumed on the walls of a room with the locations following the 1-dimensional homogeneous Poisson point process (HPPP), but the blockage effect was not considered. What's more, the numerical results showed in \cite{li2021enhancing} were merely based on Monte Carlo simulation, which may pose an issue when the indoor mmWave network scales up. \cite{qin2022indoor} considered a humanity mobility model to calculate the outage probability in a 2D indoor environment and the RIS locations (assumed to be on the wall) were optimized with respect to coverage. However, the blockage and
the RIS amount showed in the simulation section was up to 2 (e.g., 1 and 2 humans as the blockages, and 1 and 2 RISs being deployed on the wall). Such results cannot provide
insight when the network scales up. In addition, \cite{qin2022indoor} assumes multi-hop existed between the transmitter and the user\footnote{The signal can be reflected more than once by the RISs before it reaches the user.}, which is unclear currently whether RIS is feasible for this functionality.\\
\indent Through \cite{zhu2020stochastic,shafique2022stochastic,zhang2021reconfigurable,kishk2020exploiting,10064007,li2021enhancing,qin2022indoor}, it is found that most of the previous work focuses on analyzing outdoor large-scale networks (e.g., cellular networks). Although there have been some investigations on indoor mmWave networks, some of the critical aspects were overlooked, such as the blockage effect, boundary effect, the user distribution for indoor environment, and how the height of the indoor environment affects the performance of the network. In this paper, we propose a more realistic model to analyze the performance of a 3D RIS-assisted indoor mmWave network in terms of the CP. We mathematically derive the relationships between the network parameters and the CP by jointly considering the path loss,  blockage effect, and boundary effect.
\textcolor{black}{
\begin{remark}
    Unlike prior works that rely on oversimplified assumptions, the proposed model explicitly incorporates several practical factors critical to RIS-assisted indoor mmWave network design. These include height variations among transmitters, users, RISs, and obstacles, which significantly influence link distances and LOS probabilities. Furthermore, the model captures realistic user clustering behavior around indoor landmarks (e.g., kiosks, desks), and represents obstacles with non-negligible physical dimensions, such as height and radius, rather than point abstractions. By jointly considering these aspects, the model offers a more accurate and deployment-aware analysis of RIS-assisted indoor mmWave networks.
\end{remark}
}
\subsection{Contributions and Organizations}
In this paper, we investigate the performance of an RIS-assisted indoor mmWave network. The main contributions are summarized below:
\begin{itemize}
    \item \textit{An accurate and practical model is proposed to describe the small-scale indoor space}. We model the indoor environment as a cuboid where EUs follow a clustered point process, and the transmitters are deployed on the ceiling, while the RISs are deployed on the four side walls. We also consider the blockage effect by modeling the obstacles in terms of size, density, and height. Furthermore, we also consider the feasible locations of obstacles and their impact on line-of-sight (LOS) probabilities calculations, given that they are not infinitesimal points. 
    \item \textit{The expression of the CP for the RIS-assisted indoor mmWave network is derived based on the tools of stochastic geometry, and order statistic}. We derive the distributions of distances and the LOS probabilities for direct and indirect links.
    We then derive the coverage probability (CP) expression based on the distributions of distances and the LOS probabilities. Monte Carlo simulation is applied to verify the accuracy of the analysis.
    \item \textit{We propose several critical insights for RIS-assisted indoor mmWave network, which can be used to improve the coverage performance in practice}. We show the variations of the CP with respect to different network parameters through numerical results. The relationships between the transmitter number, RIS number, obstacle density, cluster radius, and the CP are revealed. These relationships will serve as guidance in the future RIS-assisted indoor mmWave network deployment.
\end{itemize}
The rest of this paper is organized as follows: Section \Romannum{2} describes the system model of the RIS-assisted indoor mmWave network. Section \Romannum{3} analyzes the distance distributions in the direct and indirect links. Section \Romannum{4} characterizes the LOS probabilities of the direct and indirect links, and Section \Romannum{5} provides the CP characterization and insights. Section \Romannum{6} shows the numerical results,  Section \Romannum{7} draws the conclusion, and Section \Romannum{8} envisages the future of the RIS-assisted mmWave indoor networks. The Notations used in this paper are explained in Table \ref{tab:notation}.
\begin{table}
\caption{Table of Notations and Descriptions}
\centering
\begin{center}
{
\renewcommand{\arraystretch}{1.15}
    \begin{tabular}{ {c} | {c} }
    \hline \hline
    \textbf{Notation} & \textbf{Description}\\ 
       \hline
       $a$ & Width and length of the cuboid\\
       \hline
       $N_T$, $N_R$, $N_o$ & \makecell{Number of transmitters,\\ RISs, and obstacles}\\
       \hline
         $M_R$ & \makecell{Number of RIS elements}\\
       \hline
       $R_b$, $R_c$ & \makecell{Radius of the obstacle,\\ and cluster}\\
       \hline
       $\lambda$ & Wavelength of the transmitted signal\\
       \hline
       $e_x$, $e_y$ & Width and length of an RIS element\\
       \hline
       $h_t$, $h_u$, $h_r$, $h_{\text{min}}$, $h_{\text{max}}$  & \makecell{Height of the transmitter, EU, RIS,\\ lowest obstacle, and highest obstacle}\\
       \hline
       $\tau$ & SNR threshold\\
       \hline
       $G_t$, $G_r$, $P_{\textrm{Tx}}$, $\sigma^2_n$  & \makecell{Transmitter antenna gain, \\receiver antenna gain, \\transmit power, noise power}\\ 
       \hline
       $\widetilde{D}_{\text{tr}}$, $D_{\text{tr}}$  & \makecell{Distance between the \\associated/random BS \\and the EU}\\
       \hline
       $\widetilde{D}_{\text{tc}}$, $D_{\text{tc}}$ & \makecell{Distance between \\the associated/random BS\\ and the cluster center}\\
       \hline 
       $\widetilde{D}_{\text{ts}}$ $D_{\text{ts}}$ &\makecell{Distance between the\\ associated/random BS\\ and the associated RIS} \\
       \hline
       $\widetilde{D}_{\text{sr}}$, $D_{\text{sr}}$ &\makecell{Distance between the\\ associated/random RIS\\ and the EU} \\
       \hline 
       $\widetilde{D}_{\text{sc}}$, $D_{\text{sc}}$ &\makecell{Distance between\\ the associated/random RIS\\ and the cluster center} \\
       \hline
       $\widetilde{D}_{\text{tr,p}}$, $\widetilde{D}_{\text{tc,p}}$, $\widetilde{D}_{\text{ts,p}}$, $\widetilde{D}_{\text{sr,p}}$, $\widetilde{D}_{\text{sc,p}}$ & \makecell{Projected distances of \\$\widetilde{D}_{\text{tr}}$, $\widetilde{D}_{\text{tc}}$, $\widetilde{D}_{\text{ts}}$, $\widetilde{D}_{\text{sr}}$, $\widetilde{D}_{\text{sc}}$ on the floor}\\
       \hline
       $P_{\text{LOS,D}}$, $P_{\text{LOS,ID}}$ & \makecell{LOS probabilities of the\\ direct and indirect links} \\
       \hline
       $P_{\text{Cov,D}}$, $P_{\text{Cov,ID}}$ & \makecell{Coverage probabilities of the\\ direct and indirect links} \\
       \hline\hline
    \end{tabular}}
\end{center}
\label{tab:notation}
\end{table}
\begin{figure*}
    \centering
    \includegraphics[width=0.8\linewidth]{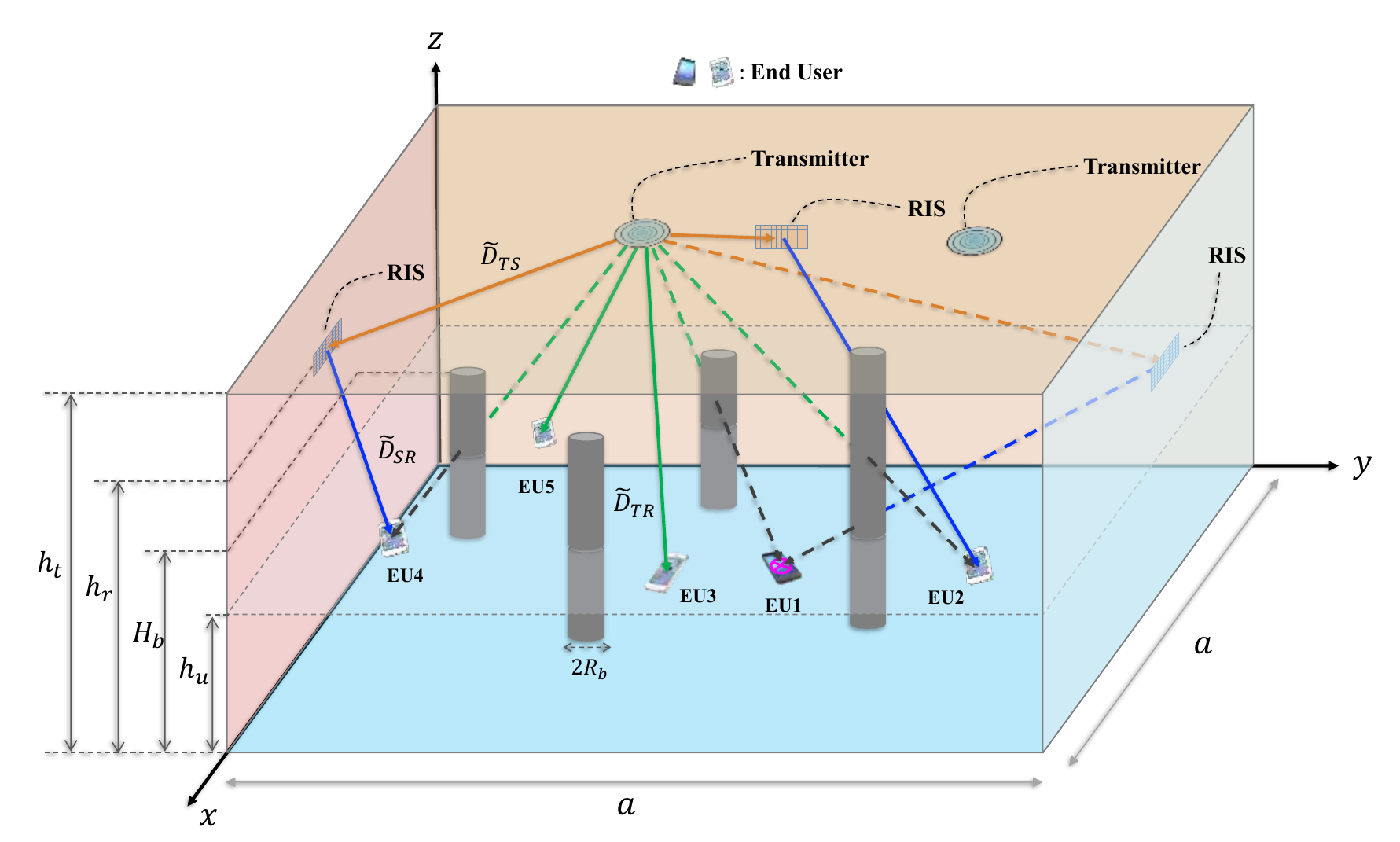}
    \caption{Illustration of the 3D RIS-assisted indoor mmWave network.}
    \label{fig:illustration_1}
\end{figure*}
\section{System Model}
 Illustration of the 3D RIS-assisted indoor mmWave network is shown in Fig. \ref{fig:illustration_1}. \textcolor{black}{In this 3D room, there are transmitters, RISs, EUs, and obstacles. Due to the blockage effects and the large-scale fading, some of the users cannot be covered (e.g., EU1 in Fig. \ref{fig:illustration_1}), while others can only be covered by the indirect link (e.g., EU2 and EU4 in Fig. \ref{fig:illustration_1}) or the direct link (EU3 and EU5 in Fig. \ref{fig:illustration_1}). To analyze how well a user is covered by this network, CP is used. Next, we introduce the system model and the essential components for the CP calculation.} The room has a length and width of $a$ and a height of $h_t$ (this room can be a conference center, a hospital, a library, or a shopping mall). \textcolor{black}{There are $N_T$ transmitters that are all deployed on the ceiling following a binomial point process (BPP) model \cite{firyaguna2020performance}}. \textcolor{black}{Each transmitter is equipped with an antenna array \cite{zhu2020stochastic}.} \textcolor{black}{There are $N_c$ clusters with radius $R_c$ in the room, where the cluster centers are modelled as BPP. Cluster centers represent real-world objects such as poster panels, information desks, or kiosks. We assume the number of users in each cluster follows an independent and identically distributed (i.i.d.) Poisson distribution, reflecting the natural variability in cluster population size, while ensuring the user count remains a non-negative integer. The users' locations in each cluster also follow BPP \cite{torres2019empirical}.} For better tractability and without loss of generality, we assume that mmWave terminals with EUs are held at the height of $h_u$. The obstacles are modeled as cylinders with radius $R_b$ and height $H_b$. The spatial distribution of the obstacle follows BPP\footnote{Because obstacles cannot overlap with each other. They should follow a hardcore process, which captures the repulsive nature of two obstacles. However, normally the area taken by the obstacle will not exceed 20\% of the room (7.86\% in \cite{ghatak2021stochastic}, 2.36\% in \cite{li2022coverage}), the probability of two obstacles closer than $2R_b$ under this condition is very small,i.e., 0.005 with $a=100~\text{m}$, $R_b = 1.25~\text{m}$, and obstacle number $N_o = 408$. Hence, BPP is a good approximation for the obstacle spatial distribution.}. 
The height of the obstacle $H_b$ is a uniform random variable, i.e., $H_b\sim \mathcal{U}(h_{\text{min}},h_{\text{max}})$. It is reasonable to have $h_{\text{min}}\leq h_u\leq h_{\text{max}}<h_t$. \textcolor{black}{The RISs are deployed on the four side walls with $N_R$ RISs on each wall and a height of $h_r$ \cite{10817377}.} To reduce the impact of obstacles on RIS-assisted communications, we place RISs higher than obstacles, i.e., $h_r\geq h_{\text{max}}$ \cite{li2022ris}. \textcolor{black}{Because there is no dependency between the obstacle distribution and RIS distribution, all cluster centers follow BPP, user numbers in different clusters follow i.i.d. Gaussian, and users in a cluster follow BPP, we assume the spatial distribution of each RIS follows a uniform distribution $\mathcal{U}(0, a)$ on each wall.}
Next, we introduce the communication strategy and the association policy between the transmitter, RIS, and user. The communication strategy includes two steps: 1) the transmitter will choose the direct link to communicate with the user, and 2) the transmitter will choose an RIS to redirect the signal to the intended user if step 1 fails. If both steps fail, we claim this communication attempt as a failure. In the first step, the user will choose the nearest transmitter. \textcolor{black}{Because $R_c$ is much smaller than $a$, we assume the nearest transmitter to the cluster center will also be the nearest transmitter to the users in this cluster.} In the second step, \textcolor{black}{the nearest RIS to the EU will be selected as the relay RIS and reflect the signal from the transmitter to the EU. In previous related work, it is commonly assumed that the RIS nearest to the transmitter is selected, or the RIS that minimizes the sum of the two distances (transmitter-to-RIS and RIS-to-EU) is selected, or the RIS that minimizes the product of these two distances is selected \cite{kishk2020exploiting,10279507, 9500724}. However, in the proposed 3D model in this paper, the blockage effect is a dominant factor compared to path loss, and there is no blockage between the transmitter and the RIS as $h_r > h_{\text{max}}$, whereas there is a possibility of blockage between the RIS and the EU. Therefore, we prioritize minimizing the distance between the RIS and the EU to achieve the lowest NLOS probability between the RIS and the EU.}\\
\indent It is widely recognized that the mmWave transmitter generally has an antenna array that can apply beamforming to users in a cluster.
In this paper, we do not consider the interference between users inside a cluster. A convincing reason is indicated in the IEEE 802.11 ad, where time division multiple access (TDMA) is applied when serving multiple users. This means even if the users are spatially close, they can still be separated in the time domain\footnote{For users that are spatially close to each other if they share the same resource (e.g., frequency, time), there will be interference because the beamforming vectors for different users will no longer be orthogonal to each other. They tend to align with each other while the spatial user correlation increases.}. Without the interference, the CP of a user is determined by the blockage effect and the large-scale fading. We define $\widetilde{D}_{\text{tr}}$ as the distance between a user and its associated transmitter. Define $\widetilde{D}_{\text{ts}}$ as the distance between the associated transmitter and the associated RIS and $\widetilde{D}_{\text{sr}}$ as the distance between the associated RIS and the user. \textcolor{black}{The CP in the proposed model comprises two components: the direct link and the indirect link. The term $P(\text{SNR}_\text{d}\geq \tau)$ represents the probability that the SNR of the direct link exceeds the detection threshold $\tau$. Meanwhile, $P_{\text{LOS,D}}(\widetilde{D}_{\text{tr}})$ denotes the probability that the direct link, with distance $\widetilde{D}_{\text{tr}}$, is line-of-sight (i.e., unblocked). Therefore, the CP through the direct link is given by the product $P_{\text{Cov-D}}= P(\text{SNR}_\text{d}\geq \tau)P_{\text{LOS,D}}(\widetilde{D}_{\text{tr}})$.\\
\indent It is important to note that the indirect link does not influence the direct link analysis, as the base station always attempts to establish a connection through the direct link first. If the direct link is either blocked or its SNR falls below the required threshold, the system resorts to the indirect link. The failure of the direct link can be mathematically expressed as
\begin{equation}
    1-P_{\text{Cov-D}}=P(\text{SNR}_{\text{d}}<\tau)+P(\text{SNR}_{\text{d}}\geq\tau)P_{\text{NLOS,D}}(\widetilde{D}_{\text{tr}})\nonumber,
\end{equation}
which captures the cases where the SNR is too low or the link is NLOS. Consequently, the CP via the indirect link is given by 
\begin{align}
    &P_{\text{Cov-RIS}}=P(\text{SNR}_{\text{id}}\geq \tau)P_{\text{LOS,ID}}(\widetilde{D}_{\text{sr}})\big(P(\text{SNR}_{\text{d}}<\tau)\nonumber \\
    &+P(\text{SNR}_{\text{d}}\geq\tau)P_{\text{NLOS,D}}(\widetilde{D}_{\text{tr}})\big)\nonumber.
\end{align}
This reflects the CP that the direct link fails while the indirect path (via the RIS) offers sufficient SNR and is unblocked. By combining the CPs of direct and indirect links, the CP of a user is then given by
\begin{align}
    &P_{\text{Cov}} = P(\text{SNR}_{\text{d}}\geq \tau)P_{\text{LOS,D}}(\widetilde{D}_{\text{tr}})\nonumber \\
    &+P(\text{SNR}_{\text{id}}\geq \tau)P_{\text{LOS,ID}}(\widetilde{D}_{\text{sr}})\big(P(\text{SNR}_{\text{d}}<\tau)\nonumber \\
    &+P(\text{SNR}_{\text{d}}\geq\tau)P_{\text{NLOS,D}}(\widetilde{D}_{\text{tr}})\big),\label{eqn:general_equation}
\end{align}
where $P_{\text{Cov}}$, $\text{SNR}_{\text{d}}$, $\text{SNR}_{\text{id}}$, $P_{\text{LOS,D}}(\cdot)$, $P_{\text{NLOS,D}}(\cdot)$, $P_{\text{LOS,ID}}(\cdot)$, and $\tau$ denote the total CP contributed by the direct and indirect links, the signal-to-noise ratio (SNR) of the direct link, the SNR of the indirect link, the LOS probability of the direct link, the NLOS probability of the direct link, the LOS probability of the indirect link and the SNR threshold for the received signal to be detected successfully. Given (\ref{eqn:general_equation}), to obtain the final expression of $P_{\text{Cov}}$, $\text{SNR}_{\text{d}}$, $\text{SNR}_{\text{id}}$, $P_{\text{LOS,D}}$ and $P_{\text{LOS,ID}}$ need to be derived, where various distributions of distances are involved in the calculation. In the next section, we show the derivations of the related distributions of distances for direct and indirect links.}

\begin{figure}
    \centering
    \includegraphics[width=0.95\linewidth]{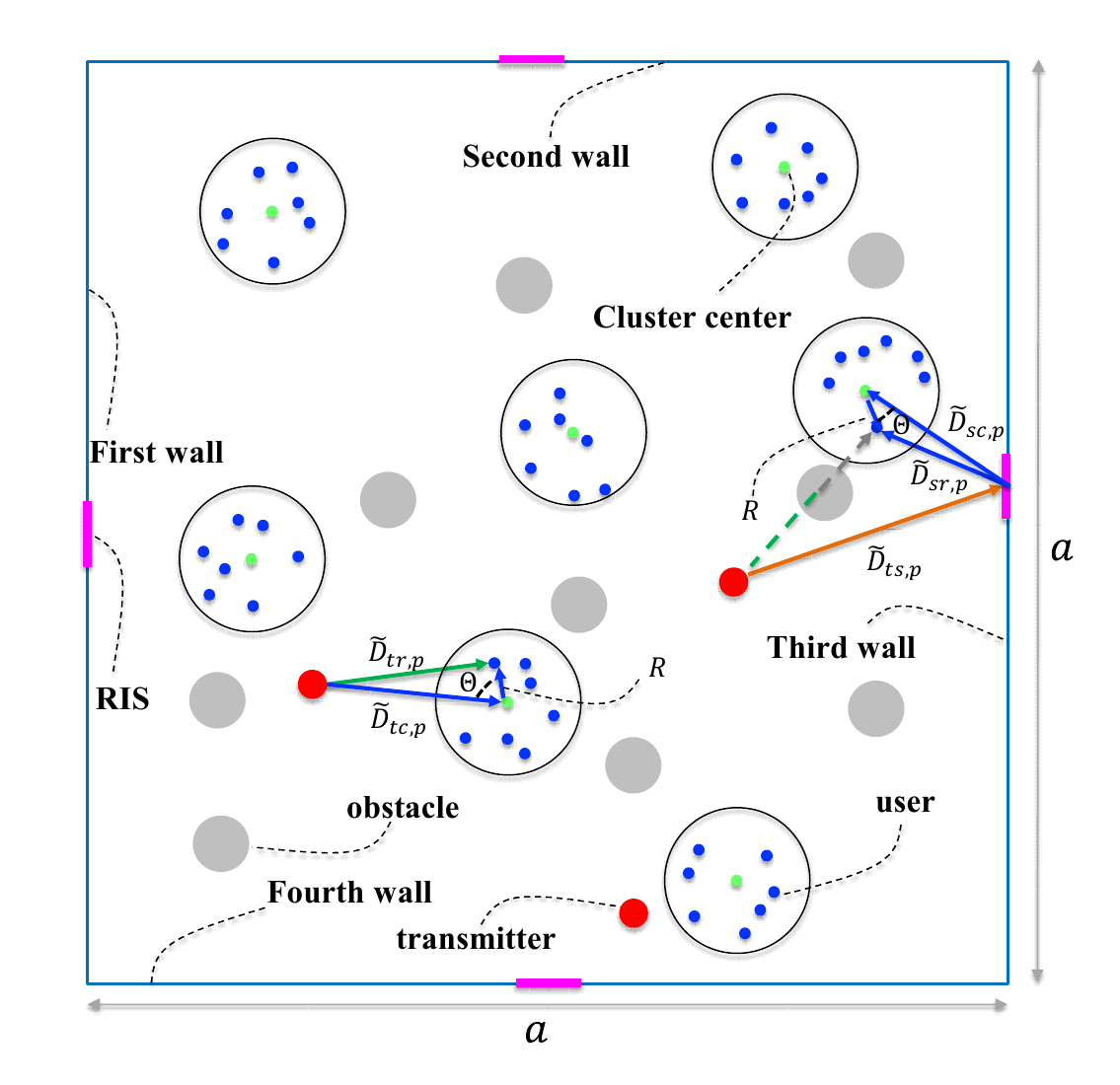}
    \caption{Illustration of the projected 3D RIS-assisted indoor mmWave network.}
    \label{fig:illustration_2}
\end{figure}
\section{Distances Distributions}
\subsection{Direct Link}
In this section, we first derive the distributions of the projected distances on the floor (shown in Fig. \ref{fig:illustration_2}) and then extend them to the 3D case. We use the tilde to represent the distance between the user and an associated node (transmitter or RIS), the symbol without a tilde represents the distance between the user and a node (transmitter or RIS). We use subscript $\text{p}$ to represent the projected lengths. We define the distance between the associated transmitter and the user in Fig. \ref{fig:illustration_2} as $\widetilde{D}_{\text{tr,p}}$ (the projected length of $\widetilde{D}_{\text{tr}}$). The square of it is given by\footnote{\textcolor{black}{Without a specific statement, we will derive the distances' squared distributions in the rest of this section for the convenience of the later sections.}}
\begin{equation}
    \widetilde{D}_{\text{tr,p}}^2 = \widetilde{D}_{\text{tc,p}}^2+R^2-2\widetilde{D}_{\text{tc,p}}R\cos{\Theta},
\end{equation}
where $\widetilde{D}_{\text{tc,p}}$, $R$, and $\Theta$ denote the distance between the associated transmitter and the cluster center, the radial distance from the user to the cluster center, and the direction of the user inside the cluster. we define $D_{\text{tc,p}}$ as the distance between a random transmitter and a random cluster center. The probability density function (PDF) is given by \cite{mathai1999introduction}
\begin{align}
    f_{D_{\text{tc,p}}}(d_{\text{tc,p}}) = \frac{4d_{\text{tc,p}}}{(a-R_c)^4}\phi(d_{\text{tc,p}}),\label{eqn:d_tc_p_simple_version}
\end{align}
where $\phi(d_{\text{tc,p}})$ is given by
\begin{align}
    \phi(d_{\text{tc,p}}) = \begin{cases}
        \frac{(a-R_c)^2\pi}{2}-2(a-R_c)d_{\text{tc,p}}+\frac{d_{\text{tc,p}}^2}{2},\\ \quad \quad \quad \quad \quad \quad \quad \quad \quad 0\leq d_{\text{tc,p}}\leq a-R_c\\
        (a-R_c)^2\bigg[\sin^{-1}{(\frac{a-R_c}{d_{\text{tc,p}}})}-\cos^{-1}{(\frac{a-R_c}{d_{\text{tc,p}}})}\bigg]\\-(a-R_c)^2-\frac{d_{\text{tc,p}}^2}{2}+2(a-R_c)\\
        \times\sqrt{d_{\text{tc,p}}^2-(a-R_c)^2},\\ 
        \quad \quad \quad \quad \quad \quad \quad \quad \quad a\leq d_{\text{tc,p}}\leq \sqrt{2}(a-R_c),
    \end{cases}
    \label{eqn:phi_x}
\end{align}
it is noted that the room length used in (\ref{eqn:d_tc_p_simple_version}) and (\ref{eqn:phi_x}) is $a-R_c$ rather than $a$. This is because the cluster center can only exist in the interior of the room with distance $R_c$ to the boundary. Hence, $D_{\text{tc,p}}$ is equivalent to the distance of two independently and uniformly distributed points in a square room with length $a-R_c$.
With the PDF of $D_{\text{tc,p}}$, and leverage order statistics, the PDF of $\widetilde{D}_{\text{tc,p}}$ is given by
\begin{equation}
    f_{\widetilde{D}_{\text{tc,p}}}(\widetilde{d}_{\text{tc,p}}) = N_Tf_{D_{\text{tc,p}}}(\widetilde{d}_{\text{tc,p}})(1-F_{D_{\text{tc,p}}}(\widetilde{d}_{\text{tc,p}}))^{N_T-1},
    \label{eqn:CDF_d_tc_p}
\end{equation}
where $F_{D_{\text{tc,p}}}(x)$ denotes the cumulative distribution function (CDF) of $D_{\text{tc,p}}$. Next, we show the derivation of $R$'s PDF. Because the users are uniformly distributed inside the circular cluster, the PDF for the location of a user in polar coordinates is given by $f_{R}(r,\theta) = 1/(\pi R_c^2)$, after taking the integral with respect to $\theta$ (a uniform random variable from 0 to $2\pi$), the PDF of $R$ is given by $ f_R(r) = 2r/R_c^2$.
The PDF of $\Omega=\cos{\Theta}$ is given by $f_{\Omega}(\omega) = 1/(\pi\sqrt{1-\omega^2})$, where $\omega\in[-1,1]$ (See proof in Appendix B). With the PDFs of $\widetilde{D}_{\text{tc,p}}$, $R$, and $\cos{\Theta}$, the CDF of $\widetilde{D}_{\text{tr,p}}$ can be expressed as
\begin{align}
   & F_{\widetilde{D}^2_{\text{tr,p}}}(\widetilde{d}^2_{\text{tr,p}}) = P\big(\widetilde{D}_{\text{tc,p}}^2+R^2-2\widetilde{D}_{\text{tc,p}}R\cos{\Theta}\leq \widetilde{d}^2_{\text{tr,p}}\big)\nonumber\\
    &=P\bigg(\cos{\Theta}\geq \frac{\widetilde{D}_{\text{tc,p}}^2+R^2-\widetilde{d}^2_{\text{tr,p}}}{2\widetilde{D}_{\text{tc,p}}R} \bigg)\nonumber \\
    &=\!\!\int\!\! f_{\widetilde{D}_{\text{tc,p}}}(\widetilde{d}_{\text{tc,p}})\!\!\int\!\! f_R(r)\bigg[ 1\!\!-\!\!\frac{\sin^{-1}\!\!{\bigg(\frac{\tilde{d}_{\text{tc,p}}^2+r^2-\widetilde{d}^2_{\text{tr,p}}}{2\tilde{d}_{\text{tc,p}}r}\bigg)}\!\!+\!\!\frac{\pi}{2}}{\pi}\bigg] \text{d}r \text{d}\widetilde{d}_{\text{tc,p}}.
\end{align}
Hence, the CDF of $\widetilde{D}_{\text{tr}}^2$ can be expressed as
\begin{align}
    &F_{\widetilde{D}_{\text{tr}}^2}(\widetilde{d}_{\text{tr}}^2) = \int f_{\widetilde{D}_{\text{tc,p}}}(\widetilde{d}_{\text{tc,p}})\int f_R(r)\nonumber \\
    &\times\bigg[ 1\!\!-\!\!\frac{1}{\pi}\sin^{-1}\!\!{\bigg(\frac{\tilde{d}_{\text{tc,p}}^2+r^2+(h_t-h_u)^2-\widetilde{d}_{\text{tr}}^2}{2\tilde{d}_{\text{tc,p}}r}\bigg)}\!\!-\!\!\frac{1}{2}\bigg] \!\text{d}r \text{d}\widetilde{d}_{\text{tc,p}}.
    \label{eqn:CDF_d_tr_square}
\end{align}
\subsection{Indirect Link}
As stated in the system model section, the associated RIS should be the closest one to the user. This means the path loss for the RIS-EU link should be minimized by the associated RIS. The path loss is proportional to the square of the distance product for the indirect link in the far-field, while the path loss is proportional to the square of the distance summation for the indirect link in the near-field \cite{ozdogan2019intelligent,tang2020wireless}. \textcolor{black}{To decide whether the communication is in the far-field or near-field, we leverage the Rayleigh distances, i.e., $Z=2D^2/\lambda$, where $D$ is the dimension of the RIS and $\lambda$ is the wavelength of the impinging waves\footnote{Rayleigh distance indicates the boundary between the near-field and far-field. If the communication distance is smaller than the Rayleigh distance, it is in the near-field, otherwise, it is in the far-field.}. Table \Romannum{2} lists various RIS dimensions, operating frequency (wavelength), and their corresponding Rayleigh distance, assuming the element spacing is $\lambda/4$ \textcolor{black}{\cite{9020088, pei2021ris}}. It is observed in Table \Romannum{2} that both the near-field and far-field exist in the indoor scenarios.
Given the general indoor application scenarios considered in this work (with dimensions of 50 m × 50 m), compared with the Rayleigh distance of 12.5 m, corresponding to a $100\times 100$ RIS operating at 30 GHz, the far-field region is the major application scenario. Moreover, smaller RIS dimensions or higher operating frequencies result in even shorter Rayleigh distances, as verified in Table \Romannum{2}.
Meanwhile, we acknowledge that the presence of a near-field region in indoor environments is inevitable. However, our analysis can be readily adapted by substituting the path-loss and RIS gain models while leaving the remainder of the framework unchanged.}\\
\begin{table}
\caption{\textcolor{black}{Rayleigh distances with different $D$ and $\lambda$}}
\centering
\begin{center}
{
\renewcommand{\arraystretch}{1.15}
    \begin{tabular}{ {c} | {c} | {c} }
    \hline \hline
    \textcolor{black}{$D$ (m)} & \textcolor{black}{$\lambda$ (mm)} & \textcolor{black}{$Z$ (m)} \\ \hline
       \hline
       \textcolor{black}{$30\lambda/4$ (30-by-30 RIS)} & \textcolor{black}{10; 6; 5} & \textcolor{black}{1.13; 0.68; 0.56}\\
       \hline
       \textcolor{black}{$40\lambda/4$ (40-by-40 RIS)}&  \textcolor{black}{10; 6; 5} & \textcolor{black}{2.00; 1.20; 1.00}\\
       \hline
      \textcolor{black}{$50\lambda/4$ (50-by-50 RIS)} & \textcolor{black}{10; 6; 5} &\textcolor{black}{3.13; 1.88; 1.56}\\ 
      \hline
      \textcolor{black}{$100\lambda/4$ (100-by-100 RIS)} & \textcolor{black}{10; 6; 5} &\textcolor{black}{12.50; 7.50; 6.25}\\ 
      \hline\hline
    \end{tabular}}
\end{center}
\label{tab:examples_far_filed_path_loss}
\end{table}
\indent We define the distance product square as $\widetilde{D}_{\text{m}}^2$. In order to find the distribution of $\widetilde{D}_{\text{m}}^2$, it is necessary to know the distribution of $\widetilde{D}_{\text{m,1}}^2$, $\widetilde{D}_{\text{m,2}}^2$, $\widetilde{D}_{\text{m,3}}^2$, $\widetilde{D}_{\text{m,4}}^2$, where $\widetilde{D}_{\text{m,1}}^2$, $\widetilde{D}_{\text{m,2}}^2$, $\widetilde{D}_{\text{m,3}}^2$, and $\widetilde{D}_{\text{m,4}}^2$ denote the distance product square on the first side wall, the second side wall, the third side wall, and the fourth side wall, because $\widetilde{D}_{\text{m}}^2 = \min\{\widetilde{D}_{\text{m,1}}^2,\widetilde{D}_{\text{m,2}}^2,\widetilde{D}_{\text{m,3}}^2,\widetilde{D}_{\text{m,4}}^2\}$. Here, we show the derivation step of $\widetilde{D}_{\text{m,1}}^2$, as others are all similar. $\widetilde{D}_{\text{m,1}}^2 = \widetilde{D}_{\text{sr,1}}^2\widetilde{D}_{\text{ts,1}}^2$, where $\widetilde{D}_{\text{sr,1}}^2$ is the distance square from the associated RIS to the user and $\widetilde{D}_{\text{ts,1}}^2$ is the distance square from the transmitter to the associated RIS. Using the law of cosine, $\widetilde{D}_{\text{sr,1}}^2$ can be expanded as
\begin{equation}
    \widetilde{D}_{\text{sr,1}}^2 = \widetilde{D}_{\text{sc,1}}^2+R^2+2R\widetilde{D}_{\text{sc,1}}\cos{\Theta},
\end{equation}
where $\widetilde{D}_{\text{sc,1}}$ is the distance from the associated RIS to the cluster center. Conditioning on the location of the cluster center $(x_c,y_c,h_u)$, the expression of $\widetilde{D}_{\text{sc,1}}$ is given by
\begin{equation}
    \widetilde{D}_{\text{sc,1}} = \sqrt{y_c^2+\widetilde{L}^2+(h_r-h_u)^2},
\end{equation}
where $\widetilde{L}$ is the distance from the nearest RIS to $(x_c,0,h_r)$. The CDF of $\widetilde{L}$ is given by
\begin{equation}
    F_{\widetilde{L}}(l) = 1-(1-F_{L}(l))^{N_R},
    \label{eqn:CDF_til_L}
\end{equation}
where $L$ denotes the distance from an RIS to $y_r$ and its CDF is given by
\begin{equation}
    F_{L}(l) = \frac{1}{a}\bigg(\min\{a,x_c+l\}-\max\{0,x_c-l\}\bigg).
\end{equation}
Hence, the CDF of $\widetilde{D}_{\text{sc,1}}$ conditioning on the location of the cluster center can be expressed as
\begin{align}
    &F_{\widetilde{D}_{\text{sc,1}}|(x_c,y_c,h_u)}(\widetilde{d}_{\text{sc,1}}) =P(\sqrt{y_c^2+\widetilde{L}^2+(h_r-h_u)^2}\leq \widetilde{d}_{\text{sc,1}})\nonumber\\
    &=P\bigg(\widetilde{L}\leq \sqrt{\widetilde{d}_{\text{sc,1}}^2-(h_r-h_u)^2-y_c^2}\bigg)\nonumber \\
    &\times\mathbf{1}(\widetilde{d}_{\text{sc,1}}^2,(h_r-h_u)^2+y_c^2),
\end{align}
where $\mathbf{1}(\cdot,\cdot)$ denotes the indicator function. After taking the integral with respect to the cluster center's location, the CDF of $\widetilde{D}_{\text{sc,1}}$ is reformulated as 

\begin{align}
&F_{\widetilde{D}_{\text{sc,1}}}(\widetilde{d}_{\text{sc,1}}) = \int_{0}^{a-2R_c}f_{X_c}(x_c)\int_{0}^{a-2R_c}f_{Y_c}(y_c)\nonumber \\
&\times F_{\widetilde{L}}\bigg(\sqrt{\widetilde{d}_{\text{sc,1}}^2-(h_r-h_u)^2-y_c^2}\bigg)\nonumber\\
&\times\mathbf{1}(\widetilde{d}_{\text{sc,1}}^2,(h_r-h_u)^2+y_c^2)\text{d}y_c\text{d}x_c,
\end{align}
where $f_{X_r}(x_r) = 1/(a-2R_c)$ and $f_{Y_r}(y_r) = 1/(a-2R_c)$ denote the PDFs of $X_c$ and $Y_c$. Based on the derivation of $\widetilde{D}_{\text{tr}}^2$, the CDF of $\widetilde{D}_{\text{sr,1}}^2$ is given by
\begin{align}
    &F_{\widetilde{D}_{\text{sr,1}}^2}(\widetilde{d}_{\text{sr,1}}^2) = \int f_{\widetilde{D}_{\text{sc,1}}}(\widetilde{d}_{\text{sc,1}})\int f_R(r)\nonumber \\
    &\!\!\times\!\! \bigg[1\!\!-\!\!\frac{1}{\pi}\sin^{-1}\!\!{\bigg(\frac{\tilde{d}_{\text{sc,1}}^2+r^2+(h_r-h_u)^2-\widetilde{d}_{\text{sr,1}}^2}{2\tilde{d}_{\text{sc,1}}r}\bigg)}\!\!-\!\!\frac{1}{2}\bigg]\!\text{d}r \text{d}\widetilde{d}_{\text{sc,1}}.
    \label{eqn:CDF_d_sr_1}
\end{align}
Next, we move forward to deriving the CDF of $\widetilde{D}_{\text{ts,1}}^2$. Conditioning on the locations of the \textcolor{black}{associated transmitter} $(x_t,y_t,h_t)$ and the cluster center $(x_c,y_c,h_u)$, $\widetilde{D}_{\text{ts,1}}^2$ can be expressed as
\begin{equation}
    \widetilde{D}_{\text{ts,1}}^2 = (x_t-x_c-\widetilde{L})^2+y_t^2+(h_t-h_r)^2.
\end{equation}
The CDF of $\widetilde{D}_{\text{ts,1}}^2$ conditioning on the locations of \textcolor{black}{associated transmitter} and cluster center can be described as
\begin{small}
    \begin{align}
    &F_{\widetilde{D}_{\text{ts,1}}^2|(x_c,y_c,h_u),(x_t,y_t,h_t)}(\widetilde{d}_{\text{ts,1}}^2) \nonumber\\
    &= P\big((x_t-x_c-\widetilde{L})^2+y_t^2+(h_t-h_r)^2\leq \widetilde{d}_{\text{ts,1}}^2 \big)\nonumber\\
    &=P\big(\widetilde{L}\geq x_t-x_c-\sqrt{\widetilde{d}_{\text{ts,1}}^2-y_t^2-(h_t-h_r)^2} \big)\nonumber\\
    &+P\big(\widetilde{L}\leq x_t-x_c+\sqrt{\widetilde{d}_{\text{ts,1}}^2-y_t^2-(h_t-h_r)^2} \big)\nonumber\\
    & = F_{\widetilde{L}}\big(x_t\!\!-\!\!x_c\!\!+\!\!\sqrt{\widetilde{d}_{\text{ts,1}}^2-y_t^2-(h_t-h_r)^2}\big)\mathbf{1}(\widetilde{d}_{\text{ts,1}}^2,y_t^2\!\!+\!\!(h_t-h_r)^2)\nonumber\\
    &-F_{\widetilde{L}}\big(x_t\!\!-\!\!x_c\!\!-\!\!\sqrt{\widetilde{d}_{\text{ts,1}}^2-y_t^2-(h_t-h_r)^2}\big)\mathbf{1}(\widetilde{d}_{\text{ts,1}}^2,y_t^2\!\!+\!\!(h_t-h_r)^2).
\end{align}
\end{small}
After taking the integral with respect to the locations of the \textcolor{black}{associated transmitter} and cluster center, the CDF of $\widetilde{D}_{\text{ts,1}}^2$ is reformulated as
\begin{small}
    \begin{align}
        &F_{\widetilde{D}_{\text{ts,1}}^2}(\widetilde{d}_{\text{ts,1}}^2) = \int_{0}^{a}f_{X_t}(x_t)\int_0^a f_{Y_t}(y_t)\int_0^{a-2R_c}f_{X_c}(x_c)\nonumber\\
        &\times\bigg( F_{\widetilde{L}}\big(x_t\!\!-\!\!x_c\!\!+\!\!\sqrt{\widetilde{d}_{\text{ts,1}}^2-y_t^2-(h_t-h_r)^2}\big)\mathbf{1}(\widetilde{d}_{\text{ts,1}}^2,y_t^2\!\!+(h_t-h_r)^2)\nonumber\\
    &-F_{\widetilde{L}}\big(x_t\!\!-\!\!x_c\!\!-\!\!\sqrt{\widetilde{d}_{\text{ts,1}}^2-y_t^2-(h_t-h_r)^2}\big)\mathbf{1}(\widetilde{d}_{\text{ts,1}}^2,y_t^2\!\!+\!\!(h_t-h_r)^2)\bigg)\nonumber\\
    &\times \text{d}x_c\text{d}y_t\text{d}x_t,
    \label{eqn:CDF_d_ts_1}
    \end{align}
\end{small}\textcolor{black}{where $f_{X_t}(x_t)=1/a$, $f_{Y_t}(y_t)=1/a$ denote the PDFs of $X_t$ and $Y_t$. Here, we briefly discuss why the marginal distribution of the associated transmitter remains uniform. Intuitively thinking, the marginal distribution of the associated transmitter should be dependent on the location of the cluster center based on the association policy. With $N_T$ increasing, the probability density of $X_t$ concentrates at $X_t = x_c$, the probability density of $Y_t$ concentrates at $Y_t = y_c$. Hence, the associated transmitter's marginal distribution will no longer remain uniform when the location of the cluster center is fixed. However, in our system model, the cluster center is assumed to follow BPP. Under this assumption, the marginal distribution of the associated transmitter's location can be viewed as the average of all the marginal distributions of every possible cluster center's location. For instance, if infinite transmitters exist on the ceiling, the probability density of $X_t=x_c$ and $Y_t=y_c$ will be 1 and 0 in other places. By taking the average of every possible location of the cluster center, the marginal distribution of the associated transmitter will be uniform.} With the CDF of $\widetilde{D}_{\text{sr,1}}^2$ and $\widetilde{D}_{\text{ts,1}}^2$, the CDF of $\widetilde{D}_{\text{m,1}}^2$ can be calculated.The CDF of $\widetilde{D}_{\text{m}}^2$ can then be given by
\begin{equation}
    F_{\widetilde{D}_{\text{m}}^2}(\widetilde{d}_{\text{m}}^2) = 1-\prod_{i=1}^4(1-F_{\widetilde{D}_{\text{m,i}}^2}(\widetilde{d}_{\text{m,i}}^2)),
\end{equation}
where $F_{\widetilde{D}_{\text{m,i}}^2}(x)$ denotes the CDF of $\widetilde{D}_{\text{m,i}}^2,~i=1,2,3,4$.
\section{LOS Probability Characterization}
\subsection{Direct Link LOS Probability Calculation}
\begin{figure}
    \centering
    \includegraphics[width=0.7\linewidth]{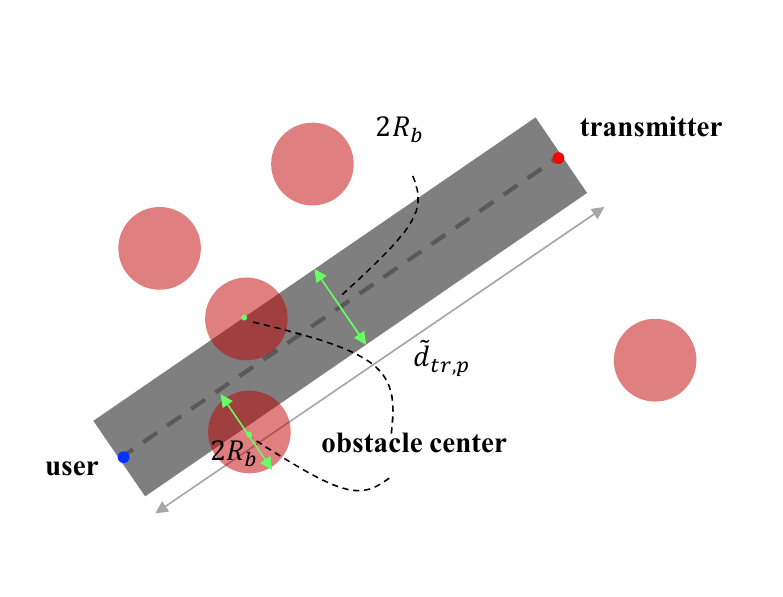}
    \caption{Illustration of the first category of the LOS probability calculation (top view).}
    \label{fig:LOS_first_category}
\end{figure}

\begin{figure}
    \centering
    \includegraphics[width=0.8\linewidth]{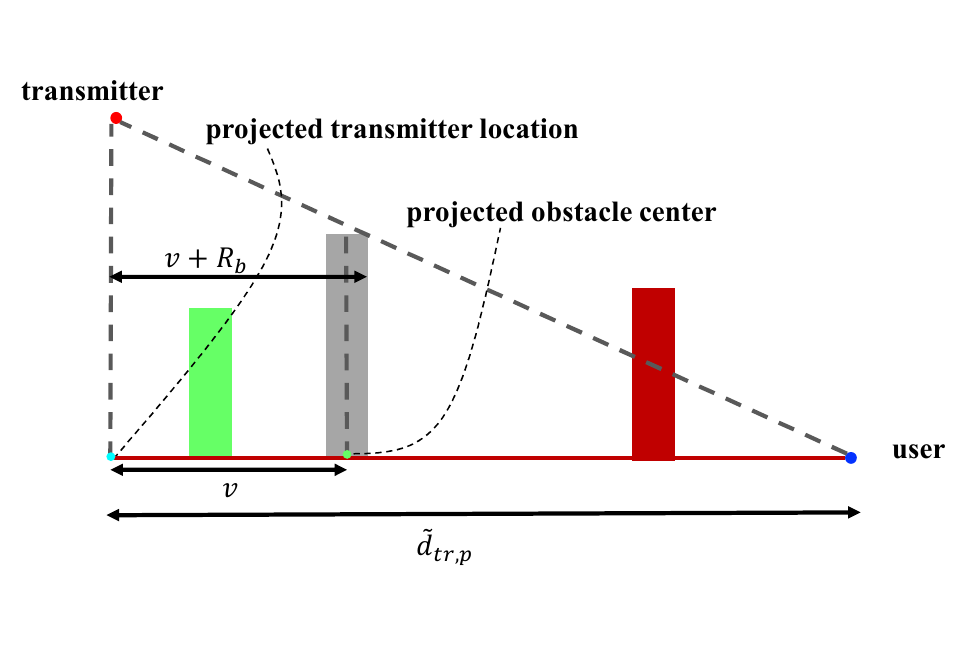}
    \caption{Illustration of the second category of the LOS probability calculation (side view).}
    \label{fig:LOS_second_category}
\end{figure}

Because the height factor is considered in this model, the condition of having the LOS path for the direct link can be categorized as: 1) every obstacle is at least $R_b$ away from the link (top view), or 2) the distance from the obstacle center to the link is smaller than $R_b$, but none of these obstacles intercept the link in the $z$ direction (side view). The first category of the LOS path can be interpreted as there is no obstacle falling into the rectangular area shown in Fig. \ref{fig:LOS_first_category}, where the width is the diameter of the obstacle and the length is $\widetilde{D}_{\text{tr,p}}$. We define the CDF of $\widetilde{D}_{\text{tr,p}}$ as $F_{\widetilde{D}_{\text{tr,p}}}(\widetilde{d}_{\text{tr,p}})$, which can be obtained through the CDF of $\widetilde{D}_{\text{tr,p}}^2$. We define the probability of no obstacle falling into the rectangular region as $P_{\text{LOS,D,1}}$, and its expression is given by
\begin{equation}
    P_{\text{LOS,D,1}} = \int f_{\widetilde{D}_{\text{tr,p}}}(\widetilde{d}_{\text{tr,p}})\Big(1-\frac{2\tilde{d}_{\text{tr,p}}R_b}{a^2}\Big)^{N_o}\text{d}\widetilde{d}_{\text{tr,p}}.
    \label{eqn:p_los_d_1}
\end{equation}
The second category can be interpreted as there are $n$ obstacles inside the rectangular region, but none of them intercept the link in the $z$ direction. The illustration is shown in Fig. \ref{fig:LOS_second_category}. The green and grey obstacles do not intercept the link while the red one intercepts the link. We define the probability that an obstacle does not intercept the link in the $z$ direction as $P_{\text{NI,D}}$. Because of the height difference between the user and the transmitter, the location of the obstacle inside the rectangular region also affects whether it blocks the link or not. More precisely speaking, the location of the obstacle projected point on the projected-transmitter-user line segment (the red line shown in Fig. \ref{fig:LOS_second_category}) affects whether the obstacle blocks the link. Conditioning on the distance between the projected transmitter location and the projected obstacle center on the line segment\footnote{Note that there are two projections for the obstacle center. The first time, it is projected onto the 2D plane, i.e., the plane with height $h_u$. The second time, it is projected on the line segment connected by the projected transmitter location and the user.} $v$ and the length of $\tilde{d}_{\text{tr,p}}$, $P_{\text{NI,D}}$ can be expressed as
\begin{equation}
    P_{\text{NI,D}} = \frac{\eta(\tilde{d}_{\text{tr,p}},v+R_b)+h_u-h_{\text{min}}}{h_{\text{max}}-h_{\text{min}}},
\end{equation}
where $\eta(\tilde{d}_{\text{tr,p}},v)$ is given by
\begin{equation}
    \eta(\tilde{d}_{\text{tr,p}},v+R_b) = \frac{h_t-h_u}{\tilde{d}_{\text{tr,p}}}\big(\tilde{d}_{\text{tr,p}}-(v+R_b)\big),
\end{equation}
After taking the integral with respect to $\tilde{d}_{\text{tr,p}}$ and $v$, $P_{\text{NI,D}}$ can be reformulated as 
\begin{equation}
    P_{\text{NI,D}} = \int\!\! f_{\widetilde{D}_{\text{tr,p}}}(\widetilde{d}_{\text{tr,p}})\!\! \int\!\! f_{V}(v)\frac{\eta(\tilde{d}_{\text{tr,p}},v\!\!+\!\!R_b) \!\!+\!\!h_u\!\!-\!\!h_{\text{min}}}{h_{\text{max}}\!\!-\!\!h_{\text{min}}}\text{d}v\text{d}\widetilde{d}_{\text{tr,p}},
\end{equation}
\textcolor{black}{where $f_{V}(v)$ is the PDF of the distance between the projected transmitter location and the projected obstacle center on the line segment. As we analyzed previously the obstacle spatial distribution can be approximated to BPP, the random variable $V$ follows uniform distribution on the line segment, i.e., $f_V(v)=1/\tilde{d}_{\text{tr,p}}$.} We define the probability $P_{\text{LOS,D,2}}$ as there are obstacles inside the rectangular region but none of them can block the link in the $z$ direction. Its expression is given by
\begin{align}
    &P_{\text{LOS,D,2}} = \int f_{\widetilde{D}_{\text{tr,p}}}(\widetilde{d}_{\text{tr,p}})\sum_{i=1}^{N_o}\binom{N_o}{i}\Big(\frac{2P_{\text{NI,D}}\tilde{d}_{\text{tr,p}}R_b}{a^2} \Big)^{i}\nonumber \\
    &\times\Big(1-\frac{2\tilde{d}_{\text{tr,p}}R_b}{a^2}\Big)^{N_o-i}\text{d}\widetilde{d}_{\text{tr,p}}.
    \label{eqn:p_los_d_2}
\end{align}
The LOS probability for the direct link $P_{\text{LOS,D}}$ is then given by
\begin{equation}
    P_{\text{LOS,D}} = P_{\text{LOS,D,1}}+P_{\text{LOS,D,2}}.
    \label{eqn:p_los_d}
\end{equation}
\begin{figure}
    \centering
    \includegraphics[width=0.8\linewidth]{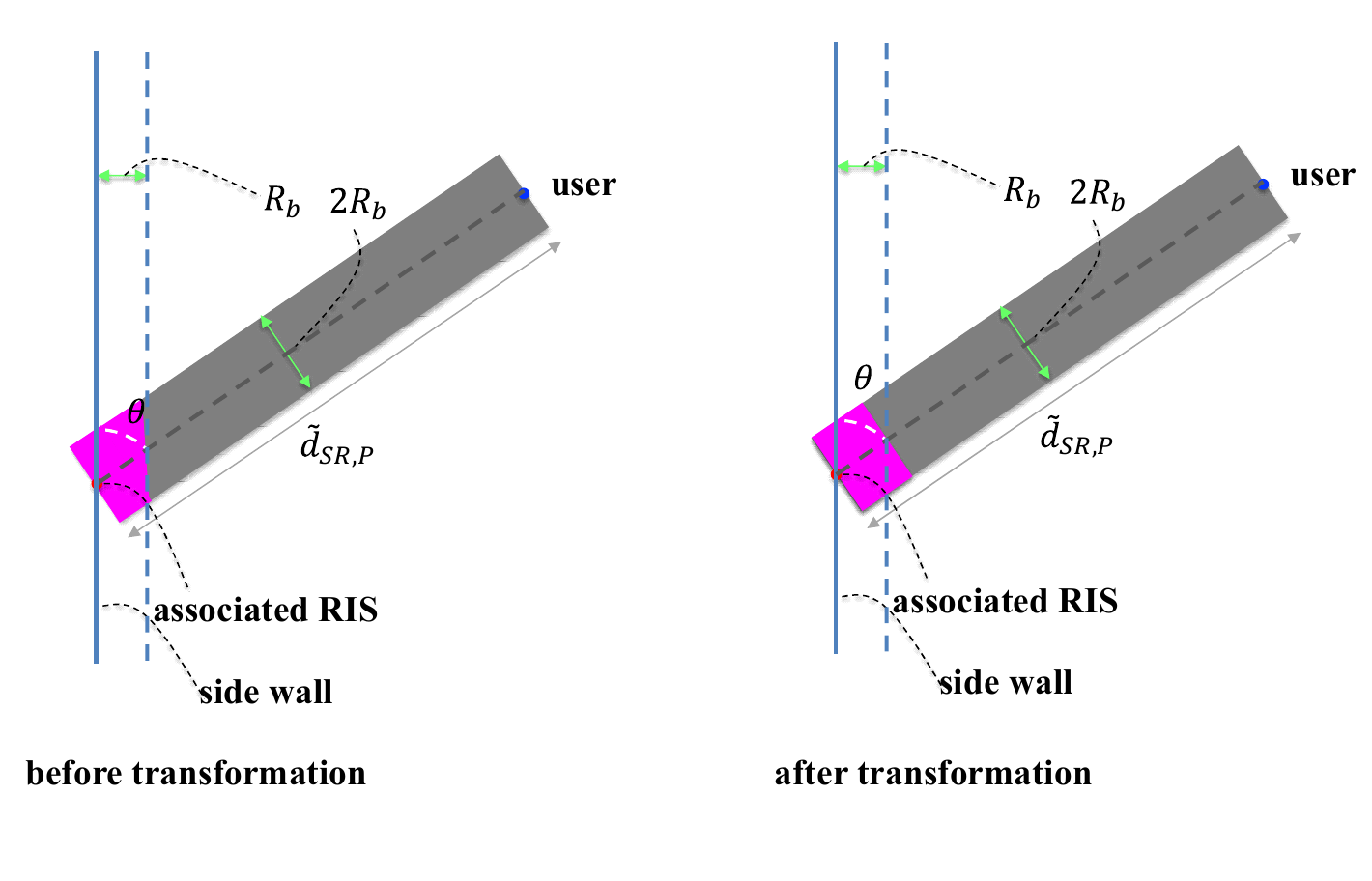}
    \caption{Illustration of the effects of boundary and feasible locations of obstacles on the indirect link LOS probability calculation.}
    \label{fig:indirect_los_illustration}
\end{figure}
\subsection{Indirect Link LOS Probability Calculation}
As we assumed $h_r\geq h_{\text{max}}$, $\widetilde{D}_{\text{ts}}$ cannot be blocked. The blockage can only happen in the associated RIS-user link. Similar to the direct link, we need the distance distribution between the associated RIS and the user, which is given by
\begin{equation}
    F_{\widetilde{D}_{\text{sr}}}(x) = 1-\prod\nolimits_{i=1}^{4}\big(1-F_{\widetilde{D}_{\text{sr,i}}}\big)
\end{equation}
where the CDF of $\widetilde{D}_{\text{sr,i}}$ can be obtained from the CDF of $\widetilde{D}_{\text{sr,i}}^2,~i=1,2,3,4$. We define the projected length of $\widetilde{D}_{\text{sr}}$ as $\widetilde{D}_{\text{sr,p}}$. As the CDF of $\widetilde{D}_{\text{sr,p}}$ can be easily obtained once knowing the CDF of $\widetilde{D}_{\text{sr}}$, we will not show the derivation of $F_{\widetilde{D}_{\text{sr,p}}}(\widetilde{d}_{\text{sr,p}})$ here. For the indirect link, there are also two categories similar to the direct link, however, the boundary effect must be taken into consideration. To better understand how the boundary affects the LOS probability for the indirect link, an illustration is given. Because the obstacle has a radius of $R_b$, it only exists on the right side of the vertical dashed line shown in Fig. \ref{fig:indirect_los_illustration}, left. This means the pink region (a right-angle trapezoid) should be deducted from the rectangular region to count the boundary effect. The area of this right-angle trapezoid is equal to the pink rectangle shown on the right side in Fig. \ref{fig:indirect_los_illustration} after the transformation. The pink rectangle area $S_r$ is given by
\begin{align}
    &S_r =\int_{\theta_1}^{\theta_2} f_{\Theta}(\theta) \frac{2R_b^2}{\sin{\theta}}\text{d}\theta=\int_{\theta_1}^{\theta_2}\frac{2}{\pi}\frac{2R_b^2}{\sin{\theta}}\text{d}\theta\nonumber\\
    &=\frac{-2R_b^2}{\pi}\bigg(\ln{\frac{1+\cos{\theta_2}}{1-\cos{\theta_2}}}-\ln{\frac{1+\cos{\theta_1}}{1-\cos{\theta_1}}} \bigg),
    \label{eqn:deducted_area}
\end{align}
where $\Theta$ is the orientation angle from $\theta_1$ to $\theta_2$. Because the variation of the pink rectangle area is symmetric in $[0,\ang{90}]$ and $[\ang{90},\ang{180}]$. We set $\theta_2 = \ang{90}$. \textcolor{black}{It is noted in (\ref{eqn:deducted_area}) that $\theta_1=\ang{0}$ will lead to an invalid expression. Hence, we set $\theta_1$ to a small value just above $\ang{0}$ (e.g., $\theta_1 = \ang{1}$). Then, (\ref{eqn:deducted_area}) can be reformulated as}
\begin{equation}
    S_r = \frac{2R_b^2}{\pi}\Big(\ln{\frac{1+\cos{\theta_1}}{1-\cos{\theta_1}}}\Big).
\end{equation}
Next, we give the expression of the LOS probability for the first category, which is
\begin{equation}
    P_{\text{LOS,ID,1}} = \int f_{\widetilde{D}_{\text{sr,p}}}(\widetilde{d}_{\text{sr,p}})\Big(1-\frac{2\tilde{d}_{\text{sr,p}}R_b-S_r}{a^2}\Big)^{N_o}\text{d}\widetilde{d}_{\text{sr,p}}.
\end{equation}
Then, we move forward to derive the LOS probability for the second category. We define the probability that an obstacle does not intercept the link in the $z$ direction as $P_{\text{NI,ID}}$. Its expression is given by 
\begin{equation}
    P_{\text{NI,ID}} = \frac{\omega(\tilde{d}_{\text{sr,p}},g+R_b)+h_u-h_{\text{min}}}{h_{\text{max}}-h_{\text{min}}},
\end{equation}
where $g$ denotes the projected obstacle location on the associated RIS-user line segment and $\omega(\tilde{d}_{\text{sr,p}},g)$ is given by
\begin{equation}
   \omega(\tilde{d}_{\text{sr,p}},g+R_b)= \frac{h_r-h_u}{\tilde{d}_{\text{sr,p}}}\big(\tilde{d}_{\text{sr,p}}-(g+R_b)\big).
\end{equation}
After taking the integral with respect to $\tilde{d}_{\text{sr,p}}$ and $g$, $P_{\text{NI,ID}}$ can be reformulated as
\begin{equation}
    P_{\text{NI,ID}} \!\!=\!\! \int f_{\widetilde{D}_{\text{sr,p}}}(\widetilde{d}_{\text{sr,p}}) \!\!\int\!\! f_{G}(g)\frac{\omega(\tilde{d}_{\text{sr,p}},g\!\!+\!\!R_b)+h_u\!\!-\!\!h_{\text{min}}}{h_{\text{max}}-h_{\text{min}}}\text{d}g\text{d}\widetilde{d}_{\text{sr,p}},
\end{equation}
where $f_{G}(g) = 1/\tilde{d}_{\text{sr,p}}$. We define the LOS probability of the second category as $P_{\text{LOS,ID,2}}$. Its expression is given by
\begin{align}
    &P_{\text{LOS,ID,2}} = \int f_{\widetilde{D}_{\text{sr,p}}}(\widetilde{d}_{\text{sr,p}})\sum_{i=1}^{N_o}\binom{N_o}{i}\Big(\frac{P_{\text{NI,ID}}(2\tilde{d}_{\text{sr,p}}R_b-S_r)}{a^2} \Big)^{i}\nonumber \\
    &\times\Big(1-\frac{2\tilde{d}_{\text{sr,p}}R_b-S_r}{a^2}\Big)^{N_o-i}\text{d}\widetilde{d}_{\text{sr,p}}.
\end{align}
The LOS probability for the indirect link $P_{\text{LOS,ID}}$ is then given by
\begin{equation}
    P_{\text{LOS,ID}} = P_{\text{LOS,ID,1}}+P_{\text{LOS,ID,2}}.
    \label{eqn:p_los_id}
\end{equation}
\textcolor{black}{
\begin{remark}
    Considering that RISs are typically deployed on the side walls of indoor environments, incorporating the boundary effect leads to a more practical and realistic system model. In bounded spaces such as conference centers or shopping malls, the LOS probability for the indirect link requires careful characterization, as the RIS is located at the boundary of the environment. 
\end{remark}
}
\section{\textcolor{black}{CP Characterization}}
\subsection{CP Characterization}
Because the interference is not considered in this paper, we use signal-to-noise ratio (SNR) to calculate the CP. The CP for the direct link is given by
\begin{equation}
    P_{\text{cov,D}} = P_{\text{LOS,D}}P\Big(\frac{G_tG_rP_{\text{Tx}}}{\sigma_n^2}\Big(\frac{\lambda}{4\pi\widetilde{D}_{\text{tr}}}\Big)^2\geq \tau\Big),
\end{equation}
    where $G_t$, $G_r$, $\sigma_n$, $\lambda$, $P_{\text{Tx}}$ and $\tau$ denote the transmitter antenna gain, the receiver antenna gain, the noise power, the wavelength of the signal, the transmit power, and the threshold. By leveraging the CDF of $\widetilde{D}_{\text{tr}}^2$, $P_{\text{Cov,D}}$ can be reformulated as
\begin{equation}
    P_{\text{Cov,D}} = P_{\text{LOS,D}}F_{\widetilde{D}_{\text{tr}}^2}\Big(\frac{G_tG_rP_{\text{Tx}}\lambda^2}{(4\pi)^2\sigma_n^2\tau} \Big).
    \label{eqn:p_cov_d}
\end{equation}
Next, we derive the CP for the indirect link. Based on \cite{ozdogan2019intelligent,tang2020wireless}, the RIS gain is proportional to the square of the RIS size. We define the total element number of an RIS as $M_R$. Define the element width as $e_x$, the element length as $e_y$. Then, the CP for the indirect link is given by 
\begin{equation}
    P_{\text{Cov,ID}} = P_{\text{LOS,ID}}P\Big(\frac{G_tG_rP_{\text{Tx}}e_xe_yM_R^2\lambda^2}{(4\pi)^2\sigma_n^2\widetilde{D}_{\text{m}}^2} \geq \tau\Big),
\end{equation}
as we have the CDF of $\widetilde{D}_{\text{m}}^2$, the CP for the indiect link can be reformulated as
\begin{equation}
    P_{\text{Cov,ID}} = P_{\text{LOS,ID}}F_{\widetilde{D}_{\text{m}}^2}\Big(\frac{G_tG_rP_{\text{Tx}}e_xe_yM_R^2\lambda^2}{(4\pi)^2\sigma_n^2\tau}\Big).
    \label{eqn:p_cov_id}
\end{equation}
Plug (\ref{eqn:p_los_d}), (\ref{eqn:p_los_id}), (\ref{eqn:p_cov_d}), and (\ref{eqn:p_cov_id}) into (\ref{eqn:general_equation}), the CP can be calculated, with the CDF of $\widetilde{D}_{\text{tr}}^2$, and $\widetilde{D}_{\text{m}}^2$ previously derived, which is given in (\ref{eqn:final_cp_exp_wto_simplification}) (on top of next page).
\begin{figure*}
    \begin{align}
&P_{\text{Cov}}\!\!=\!\!P_{\text{LOS,D}}F_{\widetilde{D}_{\text{tr}}^2}\Big(\frac{G_tG_rP_{\text{Tx}}\lambda^2}{(4\pi)^2\sigma_n^2\tau} \Big)\!\!+\!\!P_{\text{LOS,ID}}\bigg(F_{\widetilde{D}_{\text{tr}}^2}\big(\frac{G_tG_rP_{\text{Tx}}\lambda^2}{(4\pi)^2\sigma_n^2\tau} \big)P_{\text{NLOS,D}}\!\!+\!\!\Big(1\!\!-\!\!F_{\widetilde{D}_{\text{tr}}^2}\big(\frac{G_tG_rP_{\text{Tx}}\lambda^2}{(4\pi)^2\sigma_n^2\tau} \big)\Big) \bigg) F_{\widetilde{D}_{\text{m}}^2}\Big(\frac{G_tG_rP_{\text{Tx}}e_xe_yM_R^2\lambda^2}{(4\pi)^2\sigma_n^2\tau}\Big)
        \label{eqn:final_cp_exp_wto_simplification}
    \end{align}
    \hrulefill
\end{figure*}
\subsection{Relationships between Network Parameters and CP}
\begin{figure}
    \centering
    \includegraphics[width=0.8\linewidth]{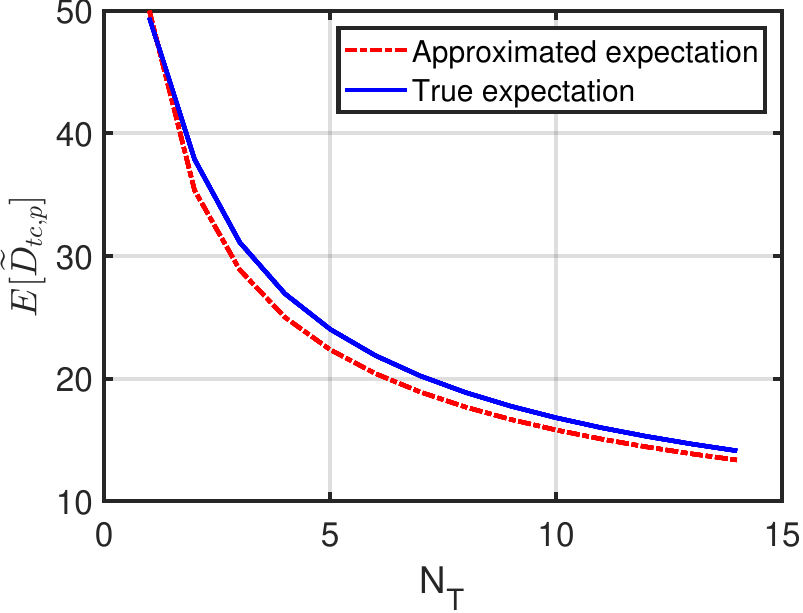}
    \caption{Comparison between the approximated and true expectation of $\widetilde{D}_{\text{tc,p}}$.}
    \label{fig:approximation_comparison}
\end{figure}

It is noted that the CP expression (\ref{eqn:final_cp_exp_wto_simplification}) is wrapped with multiple integrals, which leads to the relationships between the network parameters and the CP being difficult to extract. In this section, we reveal the relationships between network parameters (e.g., $N_T$, $N_o$, $R_c$, and $N_R$) and the CP by simplifying distance distributions (e.g., (\ref{eqn:CDF_d_tr_square}),(\ref{eqn:CDF_d_sr_1})) and expressions of LOS probabilities (e.g., (\ref{eqn:p_los_d_1}), (\ref{eqn:p_los_d_2})). These relationships will also be verified through numerical results in the next section. \\
\indent We start from (\ref{eqn:CDF_d_tr_square}) as it is a key component in (\ref{eqn:p_cov_d}). As the following inequality holds
\begin{equation}
    -1\leq \underbrace{\bigg(\frac{\tilde{d}_{\text{tc,p}}^2+r^2+(h_t-h_u)^2-\widetilde{d}_{\text{tr}}^2}{2\tilde{d}_{\text{tc,p}}r}\bigg)}_{Q}\leq 1,
\end{equation}
$\widetilde{d}_{\text{tr}}^2$ is in the range of $[(\widetilde{d}_{\text{tc,p}}-r)^2+(h_t-h_u)^2,~ (\widetilde{d}_{\text{tc,p}}+r)^2+(h_t-h_u)^2]$. We apply a Taylor series expansion around $Q=0$. Hence, $\sin^{-1}(Q)$ can be approximated as\footnote{When we vary $\widetilde{d}_{\text{tr}}^2$ conditioning on $\widetilde{d}_{\text{tc,p}}$ and $r$, $Q$ changes from 1 to -1. $Q$ will not concentrate on a specific value in the region $[-1,1]$. Hence, we choose the middle point of the region as the expansion point.}
\begin{equation}
    \sin^{-1}{(Q)}= Q+\mathcal{O}(Q^3)
    \label{eqn:tayler_expansion}
\end{equation}
as $0<|Q|< 1$, the higher order terms in (\ref{eqn:tayler_expansion}) approaches to 0.
Using this approximation and conditioning on $\widetilde{d}_{\text{tc,p}}$, (\ref{eqn:CDF_d_tr_square}) can be reformulated as
\begin{equation}
    F_{\widetilde{D}_{\text{tr}}^2}(\widetilde{d}_{\text{tr}}^2)= \frac{\widetilde{d}_{\text{tr}}^2-(h_t-h_u)^2}{\pi R_c \widetilde{d}_{\text{tc,p}}}\!-\!\frac{\widetilde{d}_{\text{tc,p}}}{\pi R_c}-\frac{R_c}{3\pi \widetilde{d}_{\text{tc,p}}}+\frac{1}{2}+\mathcal{O}(Q^3).
    \label{eqn:CDF_d_tr_square_approx_final}
\end{equation}
Through (\ref{eqn:CDF_d_tr_square_approx_final}), it is observed that the CP decreases as the cluster radius increases. As $R_c$ keeps increasing, the effect of the first term and second term in (\ref{eqn:CDF_d_tr_square_approx_final}) diminishes. Eventually, the relationship between the CP and the cluster radius is given by
\begin{equation}
    P_{\text{Cov,D}} \propto -R_c.
\end{equation}
It is also noted that in (\ref{eqn:CDF_d_tr_square_approx_final}), $N_T$ is not shown in the expression due to the conditioning on $\widetilde{d}_{\text{tc,p}}$. Next we analyze the expectation of $\widetilde{D}_{\text{tc,p}}$ to show the relationship between $N_T$ and the CP. The expectation of $\widetilde{D}_{\text{tc,p}}$ is given by
\begin{equation}
    \mathbb{E}[\widetilde{D}_{\text{tc,p}}] = \int \widetilde{d}_{\text{tc,p}}N_Tf_{D_{\text{tc,p}}}(\widetilde{d}_{\text{tc,p}})(1-F_{D_{\text{tc,p}}}(\widetilde{d}_{\text{tc,p}}))^{N_T-1}\text{d}\widetilde{d}_{\text{tc,p}}.
    \label{eqn:expectation-d_tc_p_tilde}
\end{equation}
Given the PDF expression of $D_{\text{tc,p}}$ in (\ref{eqn:d_tc_p_simple_version}) and the $N_T-1$ power term in the integrand, directly calculating (\ref{eqn:expectation-d_tc_p_tilde}) is difficult. As we know $\widetilde{D}_{\text{tc,p}}$ is the first order statistic of $D_{\text{tc,p}}$, $\widetilde{D}_{\text{tc,p}}$ is heavily influenced by the lower tail of its distribution. In other words, most of the possible values of $\widetilde{D}_{\text{tc,p}}$ will concentrate near the lower bound (0 in this context) of $D_{\text{tc,p}}$. This indicates $F_{D_{\text{tc,p}}}(\widetilde{d}_{\text{tc,p}})$ is close to 0. With $N_T$ increasing, the term $(1-F_{D_{\text{tc,p}}}(\widetilde{d}_{\text{tc,p}}))^{N_T-1}$ can be approximated as
\begin{equation}
    (1-F_{D_{\text{tc,p}}}(\widetilde{d}_{\text{tc,p}}))^{N_T-1} = e^{-N_TF_{D_{\text{tc,p}}}(\widetilde{d}_{\text{tc,p}})}.
    \label{eqn:CDF_D_tc_p_exp}
\end{equation}
The approximation comes from the limit definition of the exponential function, which is 
\begin{equation}
    \lim_{N\rightarrow\infty}(1-\frac{x}{N})^N = e^{-x}.
\end{equation}
Then, we use Taylor expansion on $F_{D_{\text{tc,p}}}(\widetilde{d}_{\text{tc,p}})$ around the point $\widetilde{d}_{\text{tc,p}}=0$.
\begin{align}
    F_{D_{\text{tc,p}}}(\widetilde{d}_{\text{tc,p}})&\approx F_{D_{\text{tc,p}}}(0)+f_{D_{\text{tc,p}}}(0)\widetilde{d}_{\text{tc,p}}+\frac{1}{2}f_{D_{\text{tc,p}}}^{\prime}(0)\widetilde{d}_{\text{tc,p}}^2\nonumber \\
    &=\frac{1}{2}f_{D_{\text{tc,p}}}^{\prime}(0)\widetilde{d}_{\text{tc,p}}^2.
    \label{eqn:CDF_taylor_d_tc_p}
\end{align}
Eq. (\ref{eqn:expectation-d_tc_p_tilde}) can be reformulated as
\begin{align}
    & \mathbb{E}[\widetilde{D}_{\text{tc,p}}] \approx N_T\int \widetilde{d}_{\text{tc,p}}f_{D_{\text{tc,p}}}(\widetilde{d}_{\text{tc,p}})e^{-N_TF_{D_{\text{tc,p}}}(\widetilde{d}_{\text{tc,p}})}\text{d}\widetilde{d}_{\text{tc,p}}\nonumber \\
     &\stackrel{(a)}{\approx}N_Tf_{D_{\text{tc,p}}}^{\prime}(0)\int \widetilde{d}_{\text{tc,p}}^2 e^{-N_TF_{D_{\text{tc,p}}}(\widetilde{d}_{\text{tc,p}})}\text{d}\widetilde{d}_{\text{tc,p}}\nonumber \\
     &\stackrel{(b)}{\approx}N_Tf_{D_{\text{tc,p}}}^{\prime}(0)\int \widetilde{d}_{\text{tc,p}}^2 e^{-\frac{N_T}{2}f_{D_{\text{tc,p}}}^{\prime}(0)\widetilde{d}_{\text{tc,p}}^2}\text{d}\widetilde{d}_{\text{tc,p}}\nonumber \\
     &=\frac{\sqrt{2\pi}}{2\sqrt{N_Tf_{D_{\text{tc,p}}}^{\prime}(0)}},
     \label{eqn:final_d_tc_pexpectation_formula}
\end{align}
where (a) comes from using Taylor expansion of $f_{D_{\text{tc,p}}}(\widetilde{d}_{\text{tc,p}})$ at point $\widetilde{d}_{\text{tc,p}}=0$, and (b) comes from (\ref{eqn:CDF_taylor_d_tc_p}). A comparison between the approximated expectation of $\widetilde{D}_{\text{tc,p}}$ and the true expectation is shown in Fig. \ref{fig:approximation_comparison} over different $N_T$. Combining (\ref{eqn:final_d_tc_pexpectation_formula}) and (\ref{eqn:CDF_d_tr_square_approx_final}), the relationship between $N_T$ and the CP is given by
\begin{equation}
    P_{\text{Cov,D}}\propto \sqrt{N_T}.
\end{equation}
Next, we show the relationship between $N_o$ and the CP. observing (\ref{eqn:p_los_d_1}) and (\ref{eqn:p_los_d_2}), the terms $(1-\frac{2\widetilde{d}_{tr,p}R_b}{a^2})^{N_o}$ and $(1-\frac{2\widetilde{d}_{tr,p}R_b}{a^2})^{N_o-i}$ can be approximated using exponential functions under large $N_o$ (similar to (\ref{eqn:CDF_D_tc_p_exp})).
\begin{align}
    &\bigg(1-\frac{2\widetilde{d}_{tr,p}R_b}{a^2}\bigg)^{N_o}= e^{-2N_o\widetilde{d}_{tr,p}R_b/a^2}\nonumber \\
    &\bigg(1-\frac{2\widetilde{d}_{tr,p}R_b}{a^2}\bigg)^{N_o-i}= e^{-2(N_o-i)\widetilde{d}_{tr,p}R_b/a^2},
\end{align}
as the distribution of $\widetilde{D}_{\text{tr,p}}$ in (\ref{eqn:p_los_d_1}) and (\ref{eqn:p_los_d_2}) will not affect the relationship between $N_o$ and the CP, it is then given by
\begin{equation}
    P_{\text{Cov,D}}\propto e^{-N_o}.
\end{equation}
For $N_T$, $N_o$ and $R_c$, the relationship will not change when switching to the indirect link. Hence, we will not derive these relationships separately for the indirect link. For $N_R$. We will make a brief discussion here. The relationship of $N_R$ with the CP resembles $N_T$ because $\widetilde{L}$ is also the first order statistic of $L$, and the PDF of $\widetilde{L}$ resembles (\ref{eqn:CDF_d_tc_p}). Following the same procedure on deriving the approximated expectation of $\widetilde{D}_{\text{tc,p}}$, the approximated expectation of $\widetilde{D}_{\text{sc,1}}$ (\ref{eqn:CDF_d_sr_1}) can also be derived.
\subsection{\textcolor{black}{Threshold-Based Analysis of RIS Contribution}}
\textcolor{black}{
Based on the previously introduced communication strategy, the base station (BS) always prioritizes the direct link; the indirect RIS-assisted link is only used if the direct link fails. As the number of transmitters and the transmit power increase, the likelihood of successfully communicating via the direct link also increases. This leads to a critical question: at what transmitter number and power level does the contribution of the RIS become negligible? We define $P_{\text{Cov-D}}=P_{\text{Cov}}-P_{\text{Cov-RIS}}$ as the CP of the direct link. We define $\varepsilon$, $\varepsilon\in(0,1)$ as the threshold at which the contribution of RIS is negligible. Then the following inequality holds,
\begin{equation}
    P_{\text{Cov-RIS}}\leq \varepsilon.
    \label{eqn:threshold_based_initial_form}
\end{equation}
(\ref{eqn:threshold_based_initial_form}) can be reformulated as 
\begin{align}
    &P_{\text{LOS,ID}}\big(1-P_{\text{Cov-D}} \big)F_{\widetilde{D}_{\text{m}}^2}\big(\frac{G_tG_rP_{\text{Tx}}e_xe_yM_R^2\lambda^2}{(4\pi)^2\sigma_n^2\tau}\big)\leq \varepsilon\nonumber \\
    &=P_{\text{LOS,ID}}\bigg(1-P_{\text{LOS,D}}F_{\widetilde{D}^2_{\text{tr}}}\big(\frac{G_tG_rP_{\text{Tx}}\lambda^2}{(4\pi^2)\sigma_n^2\tau}\big) \bigg)\nonumber\\
    &\times \underbrace{F_{\widetilde{D}_{\text{m}}^2}\big(\frac{G_tG_rP_{\text{Tx}}e_xe_yM_R^2\lambda^2}{(4\pi)^2\sigma_n^2\tau}\big)}_{J(P_{\text{Tx}})}\leq \varepsilon.
    \label{eqn:threshold_based_second_form}
\end{align}
It is hard to observe a nice and clean relationship between $N_T$, $P_{\text{Tx}}$, and $\varepsilon$ from (\ref{eqn:threshold_based_second_form}). As $P_{\text{LOS,ID}}$ is irrelevant from $N_T$ and $P_{\text{Tx}}$, we treat $P_{\text{LOS,ID}}$ as a constant $A$. For $J(P_{\text{Tx}})$, the large RIS element number ($M_R$) will push it to one. The original inequality (\ref{eqn:threshold_based_initial_form}) now becomes 
\begin{equation}
    1-P_{\text{LOS,D}}F_{\widetilde{D}^2_{\text{tr}}}\big(\frac{G_tG_rP_{\text{Tx}}\lambda^2}{(4\pi^2)\sigma_n^2\tau}\big)\leq \frac{\varepsilon}{A}.
    \label{eqn:threshold_based_third_form}
\end{equation}
As previously discussed, one of the conditions for satisfying (\ref{eqn:threshold_based_third_form}) is a large number of transmitters (e.g., 10 or 20). Under this condition, the user tends to be located closer to a transmitter, and consequently, the LOS probability of the direct link, $P_{\text{LOS,D}}$, approaches one. Hence, (\ref{eqn:threshold_based_third_form}) can be further simplified as 
\begin{equation}
    1-F_{\widetilde{D}^2_{\text{tr}}}\big(\frac{G_tG_rP_{\text{Tx}}\lambda^2}{(4\pi^2)\sigma_n^2\tau}\big)\leq \frac{\varepsilon}{A}.
    \label{eqn:threshold_based_fourth_form}
\end{equation}
We define $G_tG_r\lambda^2/(4\pi)^2\sigma_n^2\tau$ as $B$. Using (\ref{eqn:CDF_d_tr_square_approx_final}), $F_{\widetilde{D}_{\text{tr}}^2}(BP_{\text{Tx}})$ can be reformulated as
\begin{equation}
    F_{\widetilde{D}_{\text{tr}}^2}(BP_{\text{Tx}}) \!=\! \frac{BP_{\text{Tx}}\!-\!(h_t-h_u)^2}{\pi R_c\widetilde{d}_{\text{tc,p}}}\!\!-\!\!\frac{\widetilde{d}_{\text{tc,p}}}{\pi R_c}\!-\!\frac{R_c}{3\pi\widetilde{d}_{\text{tc,p}}}\!+\!\frac{1}{2}\!+\!\mathcal{O}(Q^3).
    \label{eqn:threshold_based_fifth_form}
\end{equation}
$N_T$ can be incorporated into (\ref{eqn:threshold_based_fifth_form}) by using the expectation of $\widetilde{D}_{\text{tc,p}}$, which is given in (\ref{eqn:final_d_tc_pexpectation_formula}). We define $\sqrt{2\pi}/2\sqrt{f_{D_{\text{tc,p}}}^{\prime}(0)}$ as $C$. (\ref{eqn:threshold_based_fourth_form}) can be written as 
\begin{align}
    &1 - \frac{\sqrt{N_T}\big(BP_{\text{Tx}}-(h_t-h_u)^2\big)}{\pi R_c C}+\frac{C}{\pi R_c\sqrt{N_T}}\nonumber\\
    &+\frac{R_c\sqrt{N_T}}{3\pi C}-\frac{1}{2}-\mathcal{O}(Q^3)\leq \frac{\varepsilon}{A}.
    \label{eqn:threshold_based_sixth_form}
\end{align}
By doing some transformations on (\ref{eqn:threshold_based_sixth_form}),
it can be expressed as
\begin{align}
    &P_{\text{Tx}}\geq \frac{1}{B}\bigg[(h_t-h_u)^2+\frac{\pi R_cC}{\sqrt{N_T}}\bigg(1-\frac{1}{2}-\frac{\varepsilon}{A}-\mathcal{O}(Q^3)\nonumber\\
    &+\frac{C}{\pi R_c\sqrt{N_T}}+\frac{R_c\sqrt{N_T}}{3\pi C}\bigg) \bigg].
    \label{eqn:threshold_based_seventh_form}
\end{align}
If we fix $N_T$ and treat $P_{\text{Tx}}$ as a variable, (\ref{eqn:threshold_based_seventh_form}) means as long as $P_{\text{Tx}}$ is greater than the right-hand side (RHS), the contribution of RIS to CP improvement is negligible. It can be observed that the CP improvement by RIS decreases linearly with $P_{\text{Tx}}$. Then, we fix $P_{\text{Tx}}$ and treat $N_T$ as a variable. It is observed that the RHS decreases proportionally to $1/N_T$.
}
\begin{table}
\caption{Parameter setting}
\centering
\begin{center}
{
\renewcommand{\arraystretch}{1.15}
    \begin{tabular}{ {c} | {c} }
    \hline \hline
    \textbf{Parameter} & \textbf{Value}\\ 
       \hline
       $a$ & 100 m\\
       \hline
       $N_T$, $N_R$ & [2 5 10], [1 10 20]\\
       \hline
       $N_o$ & \makecell{[21 64 107 150 193\\ 236 279 322 365 408]} \\
       \hline
       $M_R$ & 10000\\
       \hline
       $R_b$, $R_c$ & 1.25 m, 5 m\\
       \hline
       $\lambda$ & $10~\text{mm}$\\
       \hline
       $e_x$, $e_y$ & $\lambda$/2, $\lambda$/2\\
       \hline
       $h_t$, $h_u$, $h_r$, $h_{\text{min}}$, $h_{\text{max}}$  & 4 m, 1.8 m, 3.5 m, 1 m, 3 m\\
       \hline
       $\tau$ & \textcolor{black}{$1$}\\
       \hline
       \makecell{iteration time for\\ Monte-Carlo simulation} & 100000\\
       \hline
       $G_t$, $G_r$, $P_{\textrm{Tx}}$, $\sigma^2_n$  & \makecell{\textcolor{black}{100 (20 dB), 10 (10 dB)},\\ $1$ mW, $1\times 10^{-6}$ mW}\\ \hline\hline
    \end{tabular}}
\end{center}
\label{tab:validation_analytical}
\end{table}
\begin{figure*}
    \centering
    \subfigure[]{
  \begin{minipage}[b]{0.31\textwidth}
   \includegraphics[width=1.0\textwidth]{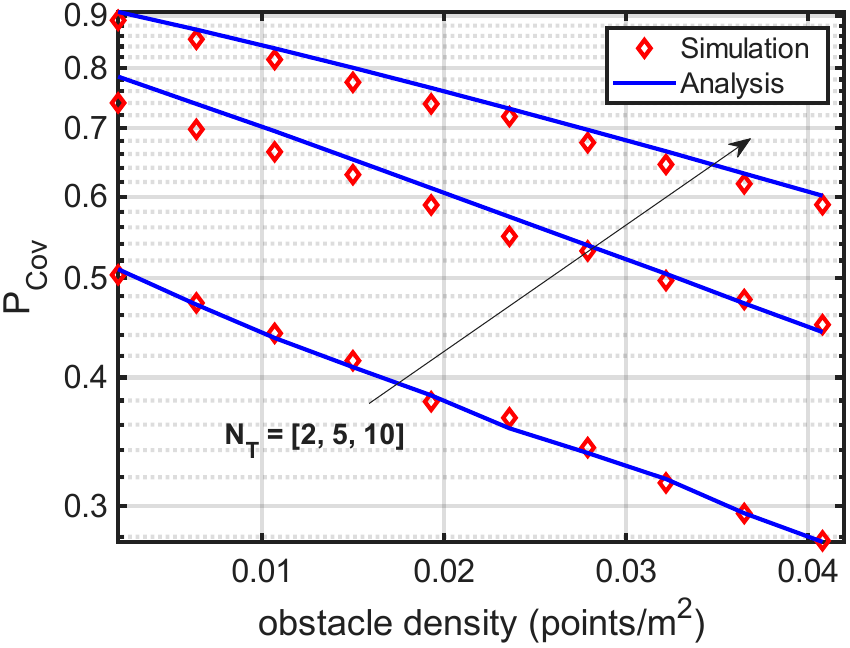}
  \end{minipage}
  \label{fig:validation_1}
 }
  \subfigure[]{
  \begin{minipage}[b]{0.31\textwidth}
   \includegraphics[width=1.0\textwidth]{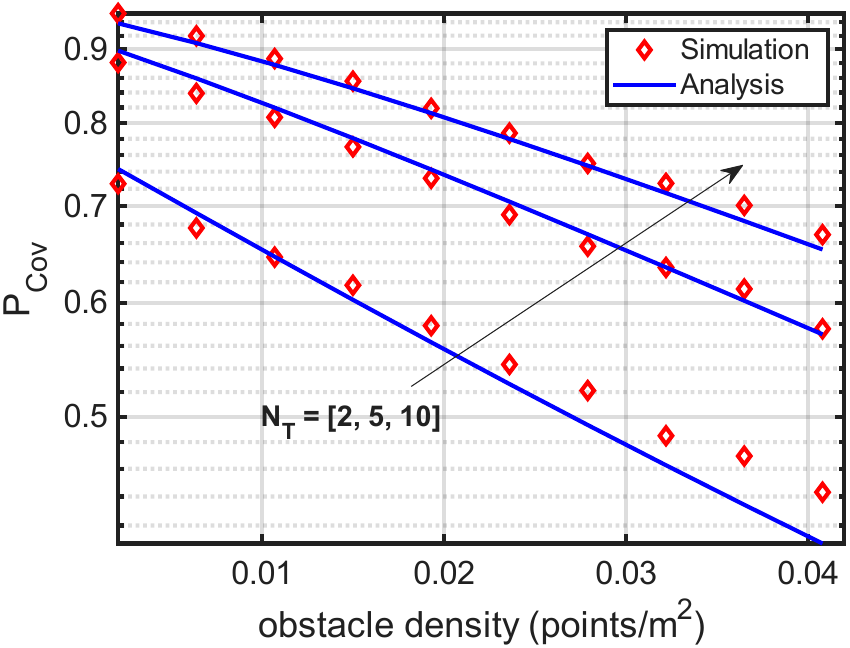}
  \end{minipage}
  \label{fig:validation_2}
 }
    \subfigure[]{
  \begin{minipage}[b]{0.31\textwidth}
   \includegraphics[width=1.0\textwidth]{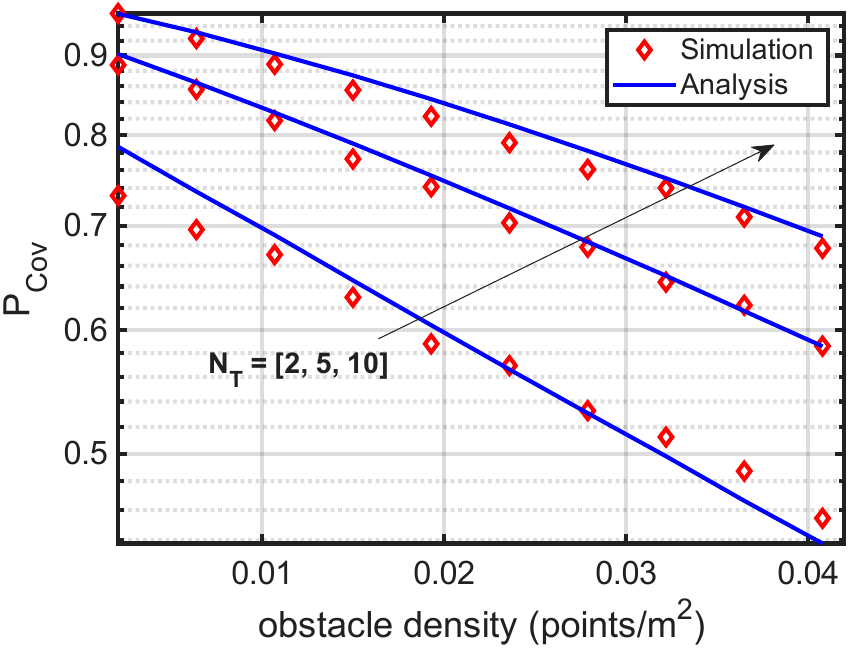}
  \end{minipage}
  \label{fig:validation_3}
 }
    \caption{(a) CP variation with respect to obstacle density with $a = 100~\text{m}$, $N_R = 1$. (b) CP variation with respect to obstacle density with $a = 100~\text{m}$, $N_R = 10$. (c) CP variation with respect to obstacle density with $a = 100~\text{m}$, $N_R = 20$.}
    \label{fig:analysis_validation}
\end{figure*}
\section{Numerical Results}
\begin{figure}
    \centering
    \includegraphics[width=0.8\linewidth]{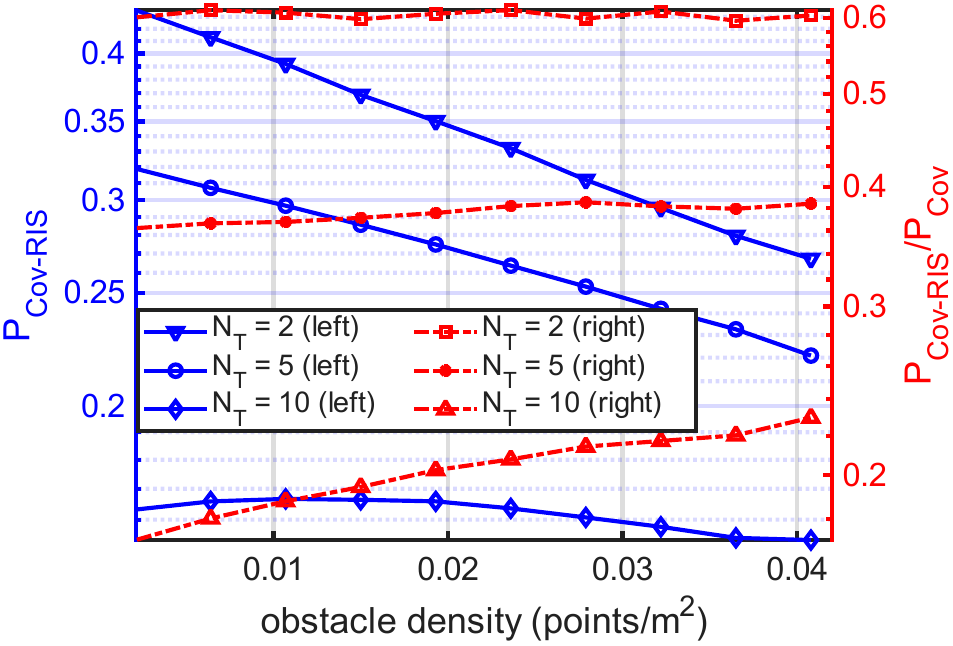}
    \caption{The contribution of RIS in terms of CP with respect to the obstacle density with $a = 100~\text{m}$, $N_R = 10$.}
    \label{fig:RIS_contribution_obstacle_density}
\end{figure}

We first validate the analytical results by comparing them with Monte Carlo simulations. The related parameter settings are given in Table \ref{tab:validation_analytical}. Observing Fig. \ref{fig:analysis_validation}, it is found that the analytical results match the Monte Carlo simulations very well under different transmitter numbers, obstacle densities, and RIS numbers. It is also found that the CP exponentially decays with the increase of the obstacle density, which agrees with the relationship given in the previous section (note that the CP is in log-scale in Fig.\ref{fig:analysis_validation}). What's more, as $N_T$ increases, the decay rate decreases, which is due to the decrease of the expected value of $\widetilde{D}_{\text{tr}}$. Now, we examine the CP variation with respect to the transmitter number. We take Fig. \ref{fig:validation_2} as an example. Observing the CP variation at a fixed obstacle density, we find that the CP increases with the increase in the transmitter number. However, the increase rate decreases with the increase in the transmitter number. For instance, the increased CP amount is around 0.2 from $N_T = 2$ to $N_T = 5$ at obstacle density =$0.02~\text{points/m}^2$, while there is only around 0.1 CP increase from $N_T = 5$ to $N_T = 10$ at the same obstacle density. Such a phenomenon agrees with what has been discussed in the previous section. \textcolor{black}{As the CP is composed of direct and indirect links, it is hard to see the contribution of RISs to the total CP. Hence, we provide Fig. \ref{fig:RIS_contribution_obstacle_density} to clearly show how the contribution of RISs varies with respect to obstacle density. Note that there are two vertical axes in Fig. \ref{fig:RIS_contribution_obstacle_density}. The left one represents the absolute CP contributed by the associated RIS (or the increased CP in the presence of RIS), and the right one represents the percentage of the RIS contribution over the total CP. It is noted that with the obstacle density increasing, the CP improved by RIS decreases. This phenomenon may seem to be counterintuitive at the beginning because RIS is designed for creating the indirect links in a rich scattering environment. In the system model, the RIS is deployed at a height exceeding that of all obstacles, ensuring that the BS-RIS link is always unblocked, regardless of obstacle density. However, an increase in obstacle density negatively impacts the LOS probability of the RIS-user link, as it becomes more susceptible to blockage. Moreover, since RISs are mounted on side walls, their placement inherently limits how close they can be to users, resulting in longer RIS-user distances on average. These two factors, reduced LOS probability and constrained deployment geometry, jointly cause the diminishing CP improvement of RIS as the obstacle density increases.
Now, we fix the obstacle density and observe the contribution of RIS under different transmitter numbers. We find that the contribution of RIS decreases with the increase in the transmitter number. Such a phenomenon can be explained in the following way: more transmitters mean the expected value of $\widetilde{D}_{\text{tr}}$ decreases, which leads to the increase in the LOS probability in the direct link. As the communication strategy stated: the transmitter will always try the direct link. It will choose an RIS for communication when the direct link fails. More transmitters indicate a low direct link failure probability. Hence, an RIS will not be selected even if the indirect link condition is feasible for communication.}\\
\begin{figure}
    \centering
    \subfigure[]{
  \begin{minipage}[b]{0.4\textwidth}
   \includegraphics[width=0.8\textwidth]{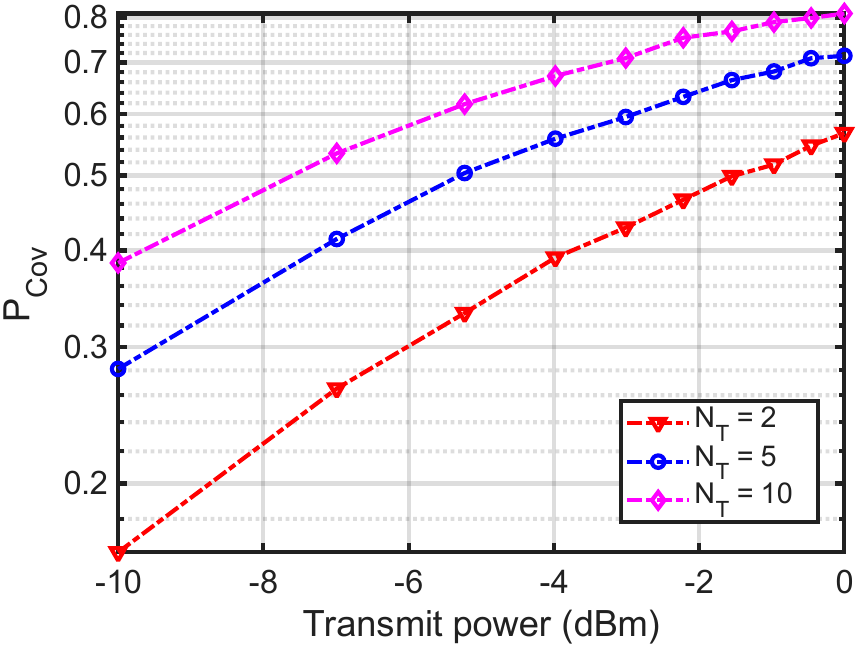}
  \end{minipage}
  \label{fig:P_tx_total}
 }
  \subfigure[]{
  \begin{minipage}[b]{0.45\textwidth}
   \hspace{0.4cm}
   \includegraphics[width=0.8\textwidth]{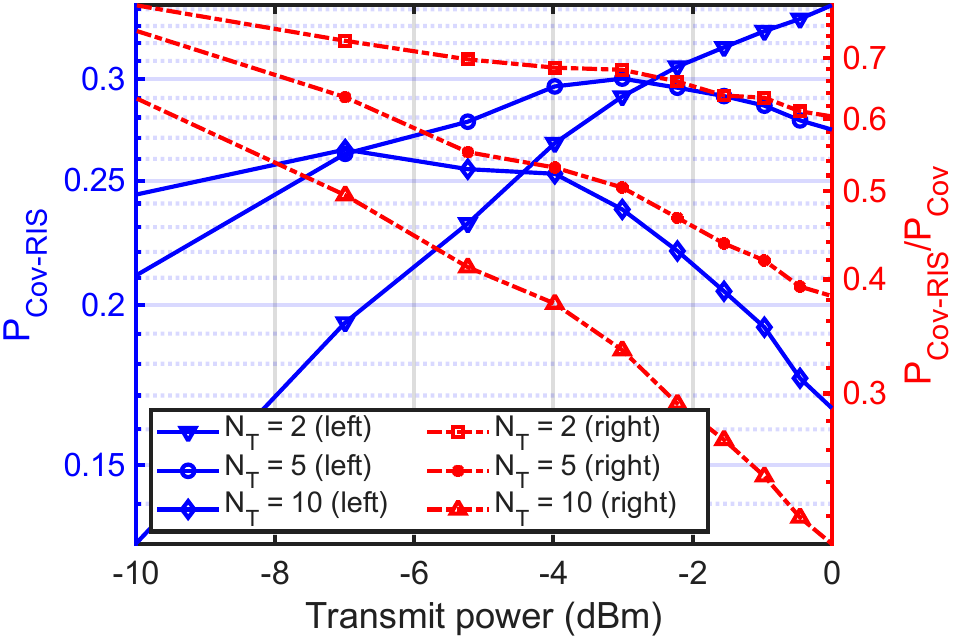}
  \end{minipage}
  \label{fig:P_tx_RIS}
 }
    \caption{(a) CP variation with respect to transmit power with $a =100~\text{m}$, $N_R = 10$, $N_o = 236$. (b) Contribution of the RIS in terms of CP with $a =100~\text{m}$, $N_R = 10$, $N_o = 236$.}
    \label{fig:P_tx}
\end{figure}
\indent Next, we show the CP variation with respect to the transmit power in Fig. \ref{fig:P_tx}. Fig. \ref{fig:P_tx_total} shows the total CP and Fig.\ref{fig:P_tx_RIS} shows the CP contributed by the RIS. Observing Fig. \ref{fig:P_tx_total}, we find that the CP increases as the transmit power increases. However, the increase amount will decrease with the increase in the number of transmitters (e.g., the increase amount is around 0.5 from -10 dBm to 0 dBm when $N_T=2$, while the increase amount is around 0.4 from -10 dBm to 0 dBm when $N_T = 10$). Observing Fig. \ref{fig:P_tx_RIS}, we find that the contribution of RIS decreases with the increase of the transmit power (right vertical axis in Fig. \ref{fig:P_tx_RIS}). However, the decrease rate varies with different transmitter numbers. The decrease rate is small at low transmitter numbers(e.g., $N_T = 2$) and large at high transmitter numbers (e.g., $N_T =5,10$). Such a phenomenon can be explained in the following way: when the transmitter number is low, the direct link failure probability is high. Hence, the contribution of RIS decreases slowly with the increase of the transmit power. With the increase of the transmitter number, the direct link failure probability decreases, and the contribution of RIS decreases fast with the increase in the transmit power. Another thing to note in Fig. \ref{fig:P_tx_RIS} is that when the transmit power is low (e.g., -10 dBm), the contribution of RIS is very high (around 80\%). This means most of the communications are assisted by the associated RIS. \\
\begin{figure}
    \centering
    \subfigure[]{
  \begin{minipage}[b]{0.4\textwidth}
   \includegraphics[width=0.8\textwidth]{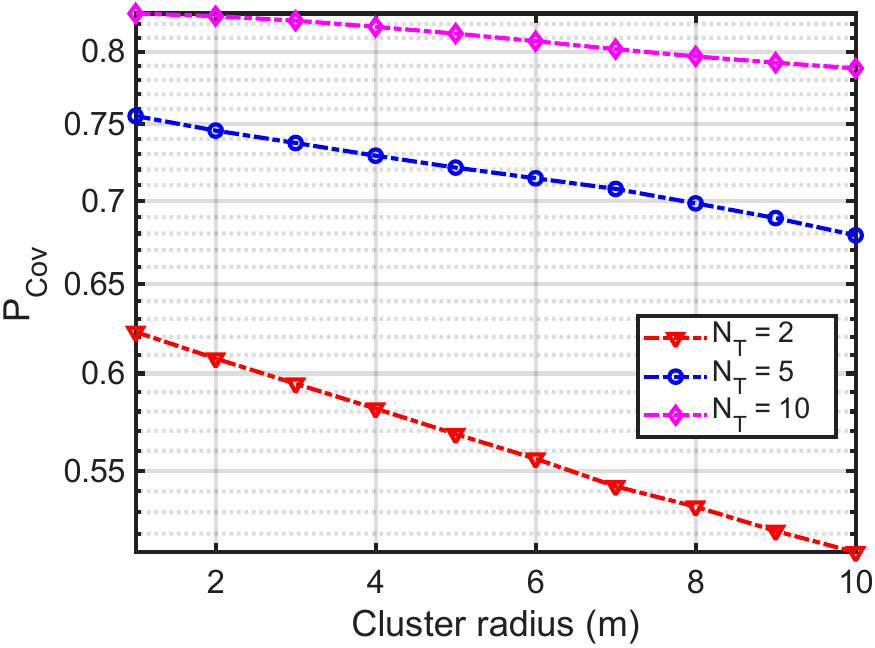}
  \end{minipage}
  \label{fig:cluster_total}
 }
  \subfigure[]{
  \begin{minipage}[b]{0.45\textwidth}
     \hspace{0.4cm}
   \includegraphics[width=0.8\textwidth]{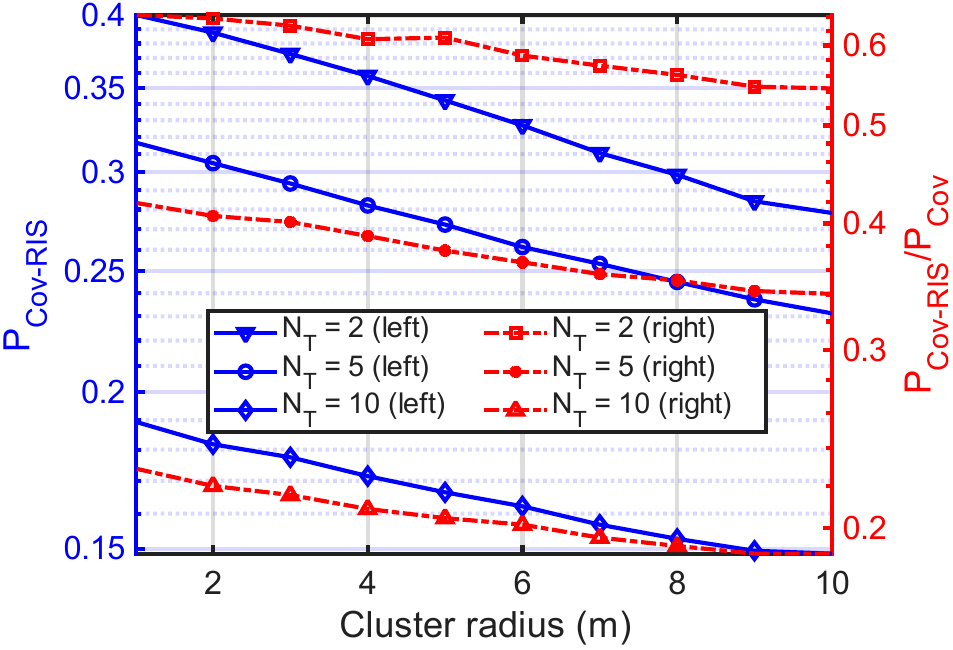}
  \end{minipage}
  \label{fig:cluster_ris}
 }
    \caption{(a) CP variation with respect to $R_c$ with $a =100~\text{m}$, $N_R = 10$, $N_o = 236$. (b) Contribution of the RIS in terms of CP with $a =100~\text{m}$, $N_R = 10$, $N_o = 236$.}
    \label{fig:cluster}
\end{figure}
\begin{figure}
    \centering
    \subfigure[]{
  \begin{minipage}[b]{0.4\textwidth}
   \includegraphics[width=0.8\textwidth]{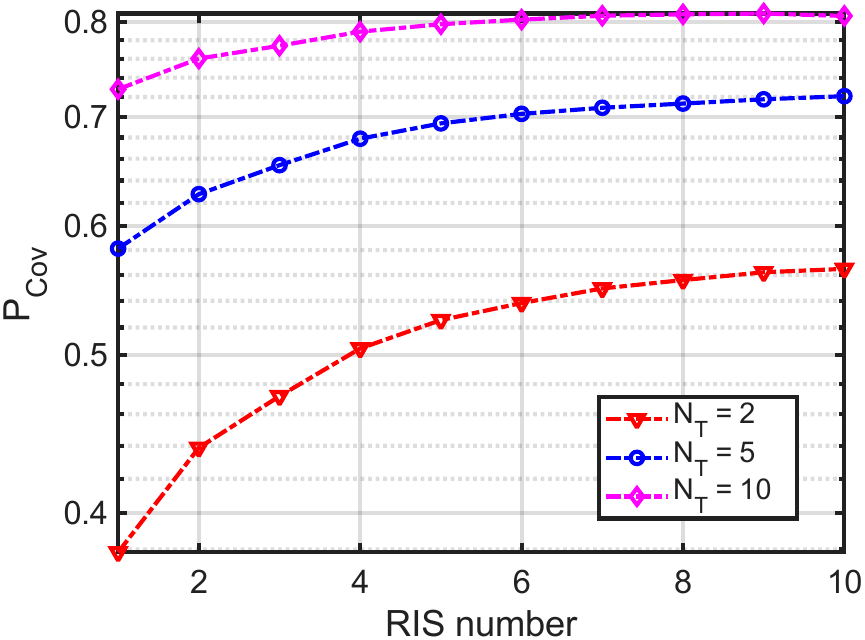}
  \end{minipage}
  \label{fig:n_r_total}
 }
  \subfigure[]{
  \begin{minipage}[b]{0.45\textwidth}
     \hspace{0.4cm}
   \includegraphics[width=0.8\textwidth]{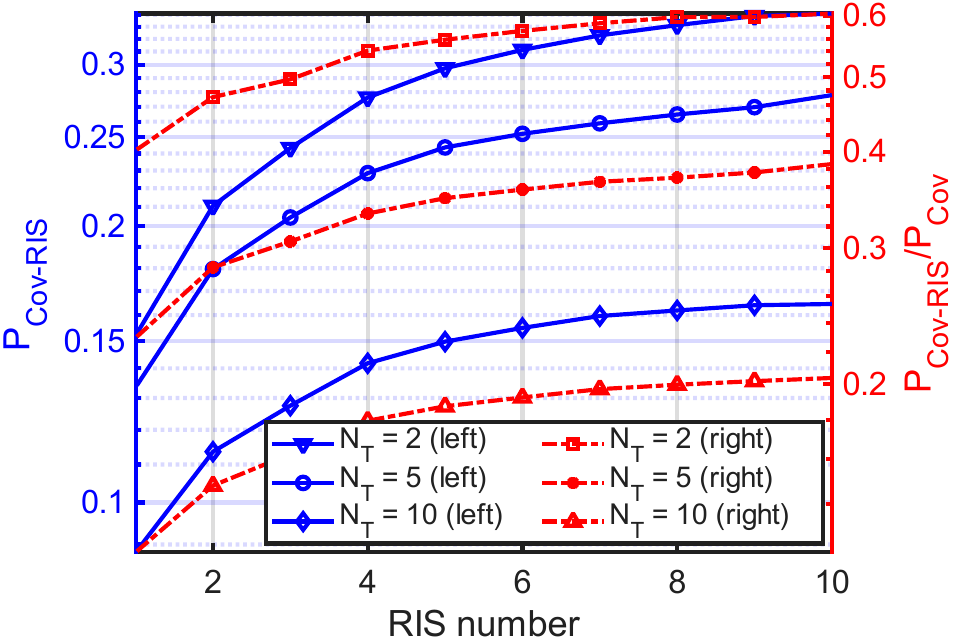}
  \end{minipage}
  \label{fig:n_r_ris}
 }
    \caption{(a) CP variation with respect to $N_R$ with $a =100~\text{m}$, $N_o = 236$. (b) Contribution of the RIS in terms of CP with $a =100~\text{m}$, $N_o = 236$.}
    \label{fig:n_r}
\end{figure}
\indent Then, we show the CP variation with respect to the cluster radius. In Fig.\ref{fig:cluster_total}, we find that the CP decreases with the increase of the cluster radius in a linear way, which matches the discussion in the previous section. However, the slope of the decrease is slower at a high transmitter number. There is around 0.05 decrease from $R_c = 1~\text{m}$ to $R_c = 10~\text{m}$ at $N_T = 10$ while there is around 0.15 decrease for $N_T = 2$. Observing Fig. \ref{fig:cluster_ris}, we find that the decrease rate of the RIS contribution does not change with respect to the transmitter number or the cluster radius (right vertical axis in Fig. \ref{fig:cluster_ris}). It is also noted in Fig. \ref{fig:cluster_ris} that the contribution of RIS is relatively high (around 40\%) at medium transmitter number (e.g., $N_T = 2,5$).\\
\indent Finally, we show the CP variation with respect to the RIS number. Through Fig. \ref{fig:n_r}, we find that the CP will increase with the increase of the RIS number at different transmitter numbers. However, the increase rate of the CP will decrease with the increase in the RIS number. This also agrees with what has been discussed in the previous section. In Fig. \ref{fig:n_r_ris}, we find that the contribution of the RIS increases with the increase of the RIS number. The contribution increase amount is roughly the same at different transmitter numbers (around 15\%). However, the absolute CP of the RIS decreases with the increase in the transmitter number. This is because the direct link failure probability decreases. \textcolor{black}{Furthermore, because the CP will eventually saturate with the increase of the RIS number, the optimal RIS number that can achieve the maximum spectral efficiency (SE) of a user can also be determined. For instance, when $N_T = 2$, the optimal RIS number for SE maximization should be 10. The optimal RIS number will decrease to 8 when $N_T=5$, and 7 when $N_T = 10$.} \\
\indent \textcolor{black}{Throughout the numerical analysis towards different network parameters, it is noted that the obstacle density is a dominant factor on the CP. With the increase of obstacle density, the CP exponentially decreases. Cluster radius, on the other hand, has a gentle effect on the CP. With the increase of the cluster radius, the CP linearly decreases. Although increasing the RIS number can mitigate the negative effects caused by the blockage and the user clustering behavior, its improvement will saturate as the RIS number increases to a certain value.\\}
\indent \textcolor{black}{
To highlight the importance of incorporating practical factors such as height and obstacle radius, we also compare the variation in CP with respect to obstacle density under two simplified models: one that neglects all nodes' heights and another that ignores the obstacle radius. The comparisons are shown in Fig. \ref{fig:comparison}. In Fig.\ref{fig:height_comparison}, it is observed that when the height of the nodes is neglected, the CP becomes significantly insensitive to the obstacle density due to the links blocked by the height of the obstacle now becoming unblocked. This simplification will overestimate the performance of the RIS-assisted indoor mmWave network. In Fig. \ref{fig:obstacle_comparison}, the comparison is made using a simplified obstacle model, where each obstacle is treated as an infinitesimal point without radius. It is observed that the CP becomes significantly less sensitive to obstacle density. This is because omitting the obstacle radius substantially increases the LOS probability for both direct and indirect links, thereby diminishing the impact of obstacle density on CP and leading to an overestimation of the network performance in terms of CP.
\begin{figure}
    \centering
    \subfigure[]{
  \begin{minipage}[b]{0.45\textwidth}
   \includegraphics[width=0.8\textwidth]{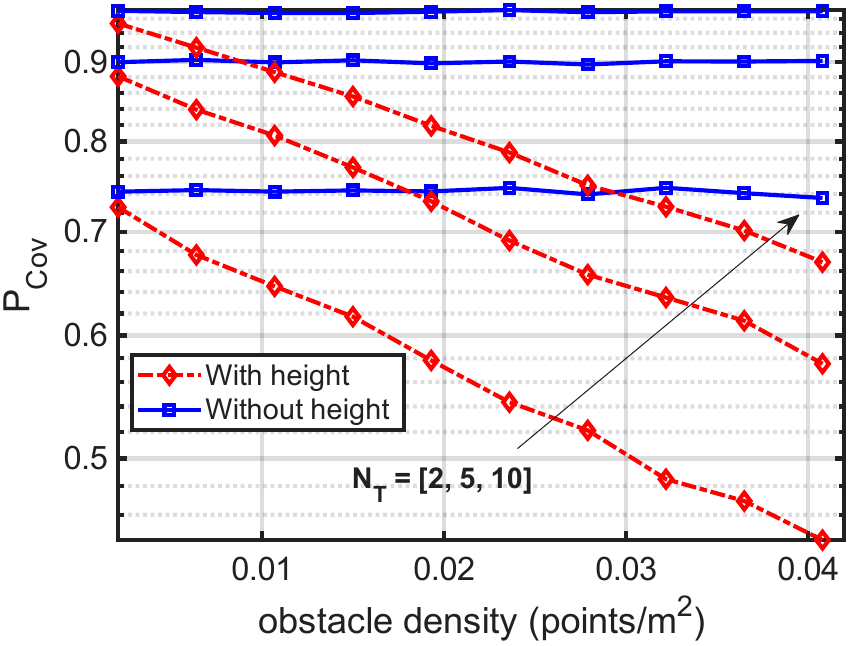}
  \end{minipage}
  \label{fig:height_comparison}
 }
  \subfigure[]{
  \begin{minipage}[b]{0.45\textwidth}
   \includegraphics[width=0.8\textwidth]{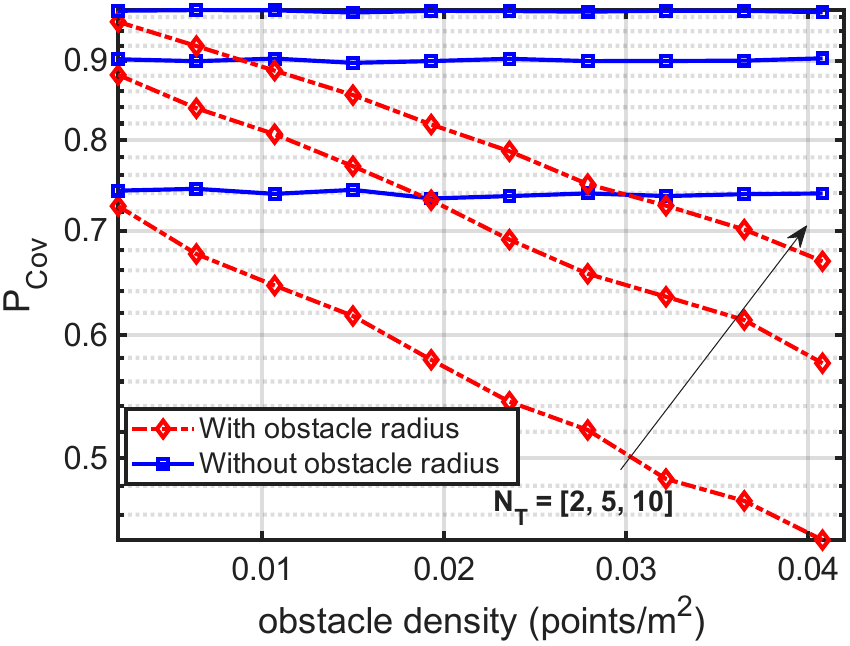}
  \end{minipage}
  \label{fig:obstacle_comparison}
 }
    \caption{(a) CP variation with respect to obstacle density when node height is omitted, for $a =100~\text{m}$, $N_R = 10$. (b) (a) CP variation with respect to obstacle density when obstacle radius is omitted, for $a =100~\text{m}$, $N_R = 10$.}
    \label{fig:comparison}
\end{figure}
}
\section{Conclusion}
In this paper, we propose an improved model and analysis for the RIS-assisted indoor mmWave network. A 3D indoor environment is modeled with the clustered user and blockage effect considered. The distance distributions related to the direct and indirect links are derived under the given communication strategy and association policy. When deriving the LOS probability for the direct and indirect links, the height factor and boundary effect are also taken into consideration. We also reveal the relationships between the network parameters and the CP. The accuracy of our analytical results is validated through comparisons with Monte Carlo simulations. Throughout the numerical results, we find that obstacle density affects the CP in a dominant way. However, this negative impact can be mitigated through the assistance of the RIS when the transmitter number is limited or the transmit power is constrained (e.g., an average of 35\% of CP improvement by RIS over different obstacle densities when $N_T=2$ and $P_{\text{Tx}}=1~\text{mW}$). Compared to obstacle density, the cluster radius has a relatively minor effect on CP and this effect will diminish as the number of transmitters increases (e.g., 15\% CP decrease from $R_c =1$ to $R_c=10$ at $N_T = 2$, while only 5\% CP decrease at $N_T=10$). It is also found that the increase of the transmitter number will diminish the contribution of RIS, which implies the RIS deployment is unnecessary in a high $N_T$ scenario, even though the environment has a high obstacle density. These relationships revealed in this paper can be used as guidance for future RIS-assisted indoor mmWave network deployment.
\section{\textcolor{black}{Road Map}}
\textcolor{black}{Although the paper proposes a more accurate and practical model for RIS-assisted indoor mmWave networks compared to existing works, several limitations remain. As the RIS dimension and operating frequency increase, the near-field region also expands. When a significant portion of the indoor environment falls within this region, it becomes necessary to characterize the system performance under near-field conditions \cite{10969557}. Moreover, since a conventional RIS can only serve one side of the space, its deployment locations are inherently restricted. A promising direction for future work is to investigate intelligent omni-surface (IOS)-aided indoor mmWave networks \cite{9722826, 11036627}, where the IOS is capable of both reflecting and refracting signals, thereby enabling full 360-degree coverage.} 
\begin{appendices}
    \section{Boundary Effect}
\begin{figure}
        \centering
        \includegraphics[width=0.65\linewidth]{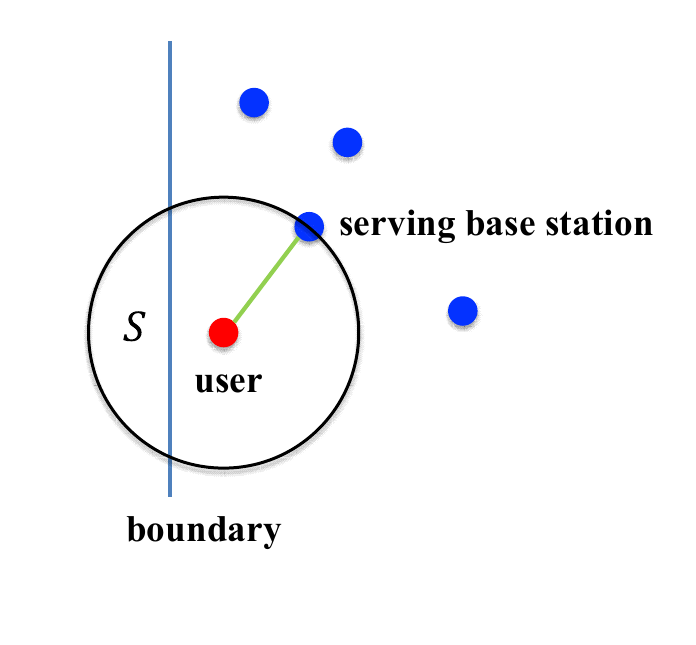}
        \caption{An illustration of the boundary effect}
        \label{fig:boundary_effect_illustration}
    \end{figure}

    The boundary effect arises when the given space is bounded (in this paper, it is a 3D indoor room) and some of the points are close to the boundary. An illustration is given to better understand this concept. In Fig. \ref{fig:boundary_effect_illustration}, a user is close to the boundary. When calculating the distance distribution from this user to its serving base station (BS). It is often assumed that the serving BS is the nearest to this user \cite{andrews2016primer}, which is equivalent to the situation that there is no BS inside the circle with radius $R$ and the user as the center. This is the conventional way to calculate the distance distribution between a user and its serving BS in outdoor networks (e.g., cellular networks). However, when the environment is bounded, such a method needs to be modified, because the equivalence of the previously mentioned two events breaks, as part of the circle ($S$ in Fig. \ref{fig:boundary_effect_illustration}) is invalid and should be deducted from the circle.
    \section{PDF derivation of $\cos{\Theta}$}
    We assume $\Omega = \cos{\Theta}$. Then, $\Theta = \cos^{-1}{\Omega}$ or $\Theta = 2\pi-\cos^{-1}{\Omega}$. The CDF of $\Omega$ is given by
    \begin{small}
        \begin{align}
        &F_\Omega(\omega) = P(\Omega\leq \omega)=P(\cos^{-1}{\omega}\leq \Theta\leq 2\pi-\cos^{-1}{\omega})\nonumber\\
        &=\frac{2\pi-2\cos^{-1}{\omega}}{2\pi}.
    \end{align}
    \end{small}
Taking the derivative of $F_{\Omega}(\omega)$ with respect to $\omega$. We can obtain
\begin{equation}
    f_{\Omega}(\omega) = \frac{1}{\pi\sqrt{1-\omega^2}}.
\end{equation}
\end{appendices}

\bibliographystyle{ieeetr}
\bibliography{reference}
\end{document}